\documentclass[usenatbib]{mn2e}
\usepackage{float}
\usepackage[dvips]{graphicx}
\usepackage{amssymb}
\usepackage{txfonts}
\usepackage{url}

\newcommand{\source}{GX 339--4}

\newcommand\mnras{MNRAS}
\newcommand\apj{ApJ}
\newcommand\aap{AA}
\newcommand\pasj{PASJ}
\newcommand\apjl{ApJ}
\newcommand\apjs{ApJS}
\newcommand\araa{ARAA}
\newcommand\aapr{A\&AR}

\newcommand\physrep{Physics Reports}
\newcommand{\msun}{{\rm M}_{\sun}}
\newcommand{\xmm}{{\textit{XMM-Newton}}}
\newcommand{\swift}{{\textit{Swift}}}
\newcommand{\nustar}{{\textit{NuSTAR}}}
\newcommand{\xte}{{\textit{RXTE}}}

\newcommand{\suzaku}{{\textit{Suzaku}}}

\title[Spectral analysis of GX 339--4 in hard state]
{Spectral analysis of the \xmm\/ data of GX 339--4 in the low/hard state: disc truncation and reflection}
\author[R. Basak and A. A. Zdziarski]
{Rupal Basak\thanks{E-mail: rupal@camk.edu.pl (RB); aaz@camk.edu.pl (AAZ)} and Andrzej A. Zdziarski$^{\star}$\\
Nicolaus Copernicus Astronomical Center, Bartycka 18, PL-00-716 Warsaw, Poland
}
\begin{document}

\date{Accepted 2016 February 19. Received 2016 February 19; in original form 2015 December 6}

\pagerange{\pageref{firstpage}--\pageref{lastpage}} \pubyear{2016}

\maketitle

\label{firstpage}

\begin{abstract}
We analyse all available observations of GX 339--4 by \textit{XMM-Newton}\/ in the hard spectral state. We jointly fit the spectral data by Comptonization and the currently best reflection code, {\tt relxill}. We consider in detail a contribution from a standard blackbody accretion disc, testing whether its inner radius can be set equal to that of the reflector. However, this leads to an unphysical behaviour of the disc truncation radius, implying the soft X-ray component is not a standard blackbody disc. This appears to be due to irradiation by the hard X-rays, which strongly dominate the total emission. We consider a large array of models, testing, e.g., the effects of the chosen energy range, of adding unblurred reflection, and assuming a lamppost geometry. We find the effects of relativistic broadening to be relatively weak in all cases. In the coronal models, we find the inner radius to be large. In the lamppost model, the inner radius is unconstrained, but when fixed to the innermost stable orbit, the height of the source is large, which also implies a weak relativistic broadening. In the former models, the inner radius correlates with the X-ray hardness ratio, which is consistent with the presence of a truncated disc turning into a complete disc in the soft state. We also find the degree of the disc ionization to anti-correlate with the hardness, leading to strong spectral broadening due to scattering of reflected photons in the reflector in the softest studied states.
\end{abstract}

\section{Introduction}
\label{intro}

Black hole X-ray binaries (BHXRB) are powered by accretion from a donor star onto a stellar-mass black hole (BH). The accretion often proceed via an optically thick, geometrically thin, accretion disc \citep{ss73,nt73}. Systems with low-mass donors usually show recurrent outbursts on a range of timescales, occurring due to a hydrogen-ionization thermal instability (e.g., \citealt*{ccl95}; \citealt{lasota01}; \citealt*{cfd12}). After a period of quiescence, an outburst starts in the low/hard spectral state, which is characterized by a power-law spectrum in the $\simeq$2--10 keV range with the photon index of $1.5\lesssim \Gamma \lesssim 2.0$ and a weak blackbody emission (e.g., \citealt*{Remillard_McClintock_2006,Doneetal_2007}). At a later phase of the outburst and at higher accretion rates, the system usually enters the high/soft spectral state \citep*{Esinetal_1997,poutanen97,Fender_Belloni_2012}, which is dominated by blackbody emission from the disc (see, e.g., \citealt{zg04} for a review of the radiation processes in different states). 

Physically, the hard state has been explained by emission from a hot inner accretion flow \citep*{SLE_1976}, surrounded by a disc that is truncated far away from the innermost stable circular orbit (ISCO). The standard hot accretion solution corresponds to the gravitational energy transferred predominantly to the ions and Coulomb energy transferred to the electrons. Given the relative weakness of the Coulomb coupling, the plasma is two-temperature with the ions being close to the virial temperature and the electrons of $k T_{\rm e}\sim 10^2$ keV. Also, the weakness of the coupling results in most of the ion energy advected to the BH at low accretion rates; thus the flow is radiatively inefficient \citep{ADAF}. However, the efficiency increases at higher accretion rates \citep{LHAF}. Furthermore, a substantial part of the accreted matter may form an outflow \citep{Blandford_Begelman_1999}. Irrespective of the detail, the outer cold disc in this scenario is expected to be truncated \citep{Doneetal_2007,gilfanov10,yn14}. The truncation is also required in early stages of an outburst, given that the disc in quiescence does not extend down to the ISCO \citep*{lny96}. When the accretion rate increases further, the truncation radius decreases, the disc starts to penetrate the inner flow and eventually reaches the ISCO in the high/soft state \citep{Remillard_McClintock_2006, Doneetal_2007}.

There is, indeed, strong observational evidence that the disc extends all the way to the ISCO in the high/soft state, e.g., \citet{Steineretal_2010}. On the other hand, the actual values of the truncation radius in the low/hard state have been a subject of an intense debate in recent years. The sources in the low/hard state usually show a soft component, which is often attributed to disc emission. The truncation radius can be calculated by modelling this component. In addition, the disc acting as a reflector produces a characteristic spectrum. The most important feature of the reflected spectrum is the fluorescent Fe K$\alpha$ line, which is broadened and skewed due to the special and general relativistic effects in regions close to the BH \citep{fabian89}. Hence, modelling the reflection component provides us another way to measure the disc truncation radius (e.g., \citealt{Reynolds_Nowak_2003}). However, various studies of the reflection spectra have resulted in conflicting results, with some studies suggesting that the accretion disc extends till the ISCO (e.g., \citealt*{miller06, rykoff07, reis10}), while some other studies find truncated discs (e.g., \citealt*{Done_Diaz_2010,plant14b,plant15}).

GX 339--4 is a BHXRB where the geometry of the low/hard state is one of the most debated. A number of studies found the disc extending down to the ISCO at a luminosity as low as $\sim$1 per cent of the Eddington luminosity, $L_{\rm E}$ \citep*{miller06, tomsick08, petrucci14}. Then, \citet{Done_Diaz_2010}, using the same data as \citet{miller06} and \citet{tomsick08}, found a large truncation radius. They argued that the discrepancy stems from the pile-up in the \xmm\/ MOS CCDs, which broadens the Fe line. Hence, one should avoid using detectors which are severely affected by this effect \citep{Done_Diaz_2010}. \citet*{kdd14} (hereafter KDD14) calculated the inner radius both from the modelling of the disc and that of the reflected component using the data taken by the \xmm\/ EPIC-pn detector in timing mode. Though their study indicates highly truncated disc, the radii calculated by the two methods do not agree with each other. \citet{plant14b} analysed the \xmm\/ EPIC-pn data taken in the imaging mode during the 2013 outburst, in which the source always remained in the hard state. They have found the inner disc radii of $R_{\rm in}\sim 20\,R_{\rm g}$ from both disc and reflection spectral components. However, the disc inclination was found to be low, $\simeq 30\degr$, which, given the large mass function of GX 339--4 found by \citet{hynes03}, would imply an unrealistically high BH mass, $M\gtrsim 40\msun$ (but see the caveats in Section \ref{gx}). On the other hand, a recent study of the same outburst using a joint \swift\/ XRT and \nustar\/ data found an inclination in the range of $\sim 40\degr$--$60\degr$ \citep{furst15}.
 
Here, we attempt to resolve some of the above controversies. For the first time, we simultaneously analyse all the available data of the source in the low/hard state taken by \xmm, whose high-resolution CCDs are crucial for accurate measurements of the reflection features. We use a large range of spectral models, in order to determine the level of robustness of our conclusions. In Section \ref{gx}, we describe main properties of GX 339--4. In Section \ref{data}, we describe the observations and data reduction, while Section \ref{analysis} describes the analysis method and the techniques that we have developed in order to get robust estimations of the disc inner radius. The results of our study are given in Section \ref{results}. In Section \ref{discussion}, we discuss physical effects related to our results and their implications. We summarize our conclusions in Section \ref{conclusions}. Appendix \ref{comparison} compares our results with those of other spectral studies.  Appendix \ref{reflection} discusses different definitions for the fractional reflection strength.

\section{The parameters of GX 339--4}
\label{gx}

The orbital period of \source\ has been measured by \citet{hynes03} as $P=1.7557$ d, though some other shorter periods were also found possible. The above period appears to be confirmed by \citet{lc06}, who found $P\simeq 1.7563 \pm 0.0003$ d by analysing the \xte\/ All Sky Monitor data. 

\citet{hynes03} measured the mass function of $f=5.8\pm 0.5\,\msun$. They used emission lines, and thus the result is less certain than that of absorption lines due to some uncertainty of the location of the emission. \citet*{munoz08} considered some of those uncertainties and argued for the mass function to be higher due to the displacement of the centre of of the emitted light with respect to the donor centre of mass, and, consequently, $M\gtrsim 7\msun$. \citet{hynes03} and \citet{munoz08} estimated the mass ratio as $M_2/M\leq 0.08$, $\leq 0.125$, respectively, where $M_2$ is the mass of the donor. 

The highest BH mass measured yet in a low-mass binary is $12.4^{+2.0}_{-1.8}\, \msun$ in GRS 1915+105 (\citealt{reid14}; see also \citealt{z14} for the dependence on the distance). Using that mass upper limit and the $1\sigma$ lower limit on $f$, the binary inclination of \source\ is $i'\geq 46\degr$. On the other hand, \citet{wu01} estimated the binary inclination to be low, $\lesssim 25\degr$, under the assumption that the velocity separation of the Balmer lines corresponds to the disc rim, though \citet{hynes03} stated that this interpretation was not certain. We note that an inner part of the disc is aligned with the BH rotation axis, which can then be inclined with respect to the binary orbit, in which case the inclination of the reflecting part of the disc, $i$, can be different than the binary one, $i'$. 

\citet{reis08} and \citet{miller08} estimated the dimensionless BH spin in \source\ as $a\simeq 0.93\pm 0.01$ while \citet*{ludlam15} claimed $a>0.97$. Those values remain, in our opinion, to be rather uncertain. We assume $a=0.9$. At this spin, the ISCO is at the radius of $r_{\rm ISCO}\simeq 2.32$ (hereafter $r$ gives the radius in the unit of $R_{\rm g}$). The actual value of $a$ has a little influence on our results because we find generally $r_{\rm in}\gg 1$, at which range the effect of the Kerr metric is small. 

The source distance is relatively uncertain, estimated at $D\simeq 8$ kpc by \citet{z04}, while \citet{hynes04} constrained it to $\gtrsim 6$ kpc. We note that \citet{z04} pointed out that the distance of $\simeq$15 kpc argued for by \citet{hynes04}, and often considered as a realistic upper limit to the distance, disagrees with the observed structure of the Galaxy. While \source\ at this distance would be $\sim 1$ kpc below the Galactic plane, \citet{hynes04} associated it with the warped material of the outer Galactic disc. However, \citet{z04} point out that the warp's Galactic azimuthal angle is $\simeq 260\degr$, while that of GX 339--4 at $D=15$ kpc would be $\simeq 320\degr$, ruling out that association. Thus, we consider $D=8\pm 2$ kpc as the currently best estimate. We note that \citet{gn06} argued for $D\sim 6$ kpc (and $M\gtrsim 10\msun$) based on comparison of the outburst light curves and Eddington ratios at state transitions with other BH transients. 

\source\ appears to be the low-mass BHXRB with the shortest firmly established recurrence time, of the order of a year to a few years. Consequently, it should have the mass transfer rate from the companion close to the the critical value (dependent on the system parameters) at which accretion becomes stable \citep{cfd12}. The mass-transfer rate can be estimated from the average flux of the system. \citet{z04} obtained $\langle F\rangle\simeq 10^{-8}$ erg s$^{-1}$ for the epoch of MJD 50800--53000. This corresponds to the average bolometric luminosity of $\langle L\rangle \simeq 8\times 10^{37}(D/8\,{\rm kpc})^2$ erg s$^{-1}$, i.e., 5 per cent of $L_{\rm E}$ of a $10\msun$ BH at 8 kpc and assuming the H fraction of $X=0.7$. The radiative efficiency, $\epsilon$, defined with respect to the mass flow through the inner Lagrangian point, $L_1$, is probably $\sim 0.1$, but it can be much smaller than that in the presence of advection and outflows, see, e.g., \citet{yn14}. We can write the corresponding mass transfer rate as $-\dot M_2=4\upi D^2\langle F\rangle/(\epsilon c^2)\simeq 1.4\times 10^{-8} (\epsilon/0.1)^{-1} (D/8\,{\rm kpc})^2\, \msun\,{\rm yr}^{-1}$. Almost the same value, $1.3\times 10^{-8} (D/8\,{\rm kpc})^2\, \msun\,{\rm yr}^{-1}$, was obtained for the epoch of MJD 50100--55900 by \citet{cfd12}. According to them, this corresponds to about 40 per cent of the critical mass transfer rate.

On the other hand, \citet{munoz08} calculated the theoretical mass transfer rate in \source\ using the model of \citet{king93} (who used the evolutionary calculations of \citealt*{webbink83}) as $-\dot M_2< 8\times 10^{-10} \, \msun\,{\rm yr}^{-1}$, i.e., $\gtrsim$16 times less than the observational estimate of \citet{cfd12}. Clearly, that theoretical maximum is strongly violated by the observational data. Thus, the results of \citet{munoz08} for the donor evolutionary status and mass based on those calculations may be not accurate.

\section{Observations and data reduction}
\label{data}
\begin{table*}\centering
\caption{Observations of GX 339--4 with the \xmm\/ EPIC-pn detector, ordered according to the decreasing X-ray hardness. The fluxes are for unabsorbed emission, the bolometric luminosity, $L$, has been calculated based on simultaneous observations by \xte\/ (for obs.\ 1, 2, 3, 7) and \swift\/ BAT (for obs.\ 4, 5, 6), see Section \ref{xte}. The Eddington ratio assumes $10\msun$, 8 kpc and $X=0.7$.}
\begin{tabular}{ccccccccc}
\hline
\# & Obs. ID & Date & Mode & Rate & Exposure & $F(0.7$--$10\,{\rm keV})$ & $F(0.01$--$10^3$\,keV) & $L/L_{\rm E}$\\
& & & & ct\,s$^{-1}$ & ks & $10^{-9}$\,erg\,cm$^{-2}$\,s$^{-1}$ & $10^{-9}$\,erg\,cm$^{-2}$\,s$^{-1}$\\
\hline
1 & 0605610201 & 2009-03-26 & Timing & 87 & 17.1 & 1.09 & 5.1 & 0.027\\
2 & 0204730301 & 2004-03-18 & Timing & 210 & 71.2 & 2.74 & 10.4 & 0.054\\
3 & 0204730201 & 2004-03-16 & Timing & 197 & 85.1 & 2.58 & 10.6 & 0.055\\
4 & 0692341301 & 2013-09-30 & Imaging & 49 & 9.2 & 1.33& 7.4 & 0.039\\
5 & 0692341401 & 2013-10-01 & Imaging & 45 & 14.7 & 1.21& 6.8 & 0.035\\
6 & 0692341201 & 2013-09-29 & Imaging & 47 & 7.3 & 1.27& 7.4 & 0.039\\
7 & 0654130401 & 2010-03-28 & Timing & 349 & 21.6 & 10.9 & 29 & 0.152\\
\hline
\end{tabular}
\label{tobs}
\end{table*}

The EPIC-pn detector on board of \xmm\/ has an energy resolution of $\simeq$2--3 per cent at 6 keV, compared, e.g., to $\sim$18 per cent at 6 keV of the \xte\/ Proportional Counter Array (PCA). Given that relatively high resolution, we have chosen to study all the currently available \xmm\/ observations in the hard state of GX 339--4. There are seven of them obtained between 2004 and 2013, see Table~\ref{tobs}. Their bolometric fluxes correspond to the Eddington ratios of $\simeq 3$--15 per cent. The order of the observations is shown in the decreasing hardness, i.e., following the usual temporal outburst evolution in the rising stage. The observations 4--6 were taken in the small window imaging  mode during the decay phase of the outburst of 2013, while the remaining data were taken in the timing mode during the rising phases of 2004 and 2009--2010 outbursts.

CCD detectors are prone to pile-up at high count rates, such as typical for Galactic BHXRBs, in particular GX 339--4. Hence, we do not include the EPIC-MOS cameras, which suffer from pile-up more than the EPIC-pn camera by a factor of $>5$, see the \xmm\/ Users Handbook. We use the Science Analysis System (SAS) version 14.0.0 for the data reduction from the raw observation data files, applying the current calibration files. For the observations taken in the imaging mode (obs.\ 4--6), we extract the data from a circular region with an aperture of  $45\arcsec$ around the source direction. For the other observations, which are in the timing mode, we choose a region in {\tt RAWX} of [31:45]. The SAS task {\tt EPATPLOT} is used to check for photon pile-up and pattern pile-up (denoting erroneous energy assigned to photons impinging on the same or neighbouring pixels). For the observations in the imaging mode, the pile-up is still high. We check its level by excluding circular regions with different radii around the source. The pile-up level converges to $<$1 per cent (single and double events) for exclusion of a circular region with a radius of $11.5\arcsec$ (see also \citealt{plant14b}). Among the timing mode data, only the observation 7 is mildly affected, which is mitigated by excluding the three middle {\tt RAWX} bins from the data. 

After that, the spectra have been extracted with the standard procedures, using only single and double events (${\tt PATTERN}\leq 4$) and ignoring bad pixels.  We also correct for flaring particle background using the standard procedure.  The SAS tools {\tt RMFGEN} and {\tt ARFGEN} are used to generate the response and the ancillary files, respectively. The spectral files, response and ancillary files are then combined using the task {\tt SPECGROUP}. We use a spectral oversampling factor of 3, which is recommended for the EPIC detectors, and a minimum of 25 counts per bin. We also include a systematic error of 1 per cent in all of the spectra. 

The imaging-mode data of observations 4--6 provide the widest energy bandwidth of 0.3--10 keV. This is important for the modelling of the disc emission. However, we find strong spectral features below 0.4\,keV (also reported by \citealt{plant14b}), which appear to be instrumental. Hence, we neglect the channels below 0.4 keV. For the data taken in the timing mode, we use the 0.7--10\,keV range, as recommended by the \xmm\/ Users Handbook. However, in both imaging and timing mode data we exclude the energy band of 1.75--2.35\,keV, which shows strong features in all of our spectra. They are apparently instrumental (and frequently present in \xmm\/ observations of bright sources).

\begin{figure}\centering{
\includegraphics[width=\columnwidth]{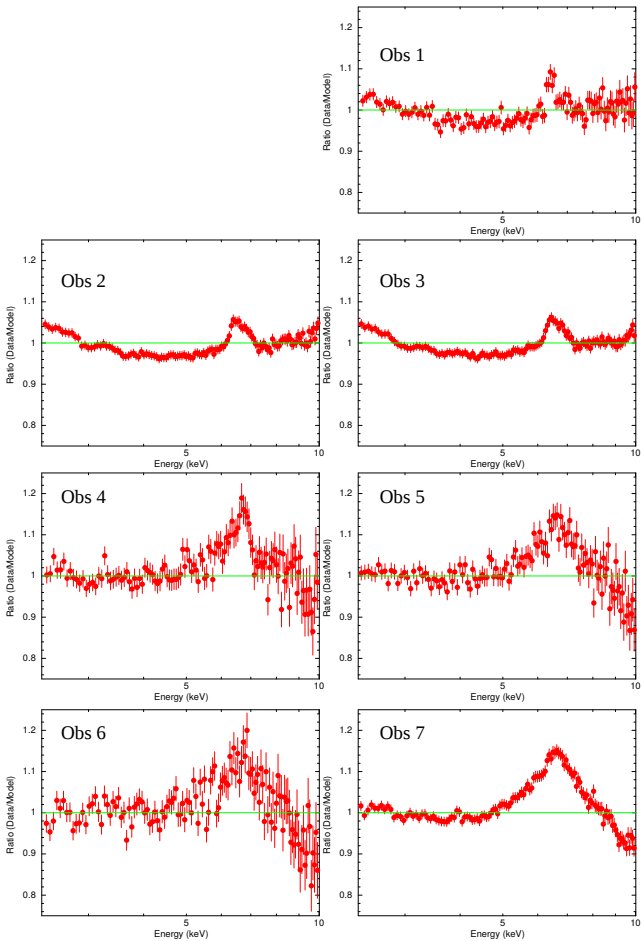} }
\caption{The profiles of the Fe K region for the studied observations, shown as the ratios of the data to the absorbed power-law model excluding the 4--7\,keV range. See Section \ref{data} for details.
}
\label{Fe_line}
\end{figure}

It is customary to characterize the broad Fe line region by fitting the data neglecting the 4--7\,keV band and then showing the ratio of data to model. In Fig.\ \ref{Fe_line}, we show such plots for all the observations. We have fitted an absorbed power law to the data in the 2.35--4 and 7--10\,keV ranges. The data at lower energies are neglected due to uncertainty in the response and the unphysical nature of the {\tt diskbb} model when applied to the hard state (see Section~\ref{cases}). As the fit is done in a narrow band, we use the parameters of the absorption component for our best case 2(i), see below. We see that the obtained line is quite narrow for obs.\ 1--3, which have the hardest spectra, but become broader as the spectrum softens, which is seen most pronouncedly for obs.\ 5--7. We see that the line width increases with the spectral softness, rather than the flux, which is lower for the obs.\ 5--6 than for the 2--3, see Table \ref{tobs}. We note that this strong increase of the line width with the softness appears already incompatible with the claims \citep{miller06,miller08,reis10} of the disc extending to the ISCO for the obs.\ 2, 3, which have the Fe lines among our narrowest. 

\section{The Analysis Method}
\label{analysis}
\subsection{The model}
\label{model}

In the hard state, the X-ray continuum at $E\gtrsim 1$ keV is dominated by thermal Comptonization. There is also a component due to reflection of this continuum from the disc (or another form of optically-thick matter). In addition, a soft excess is usually present in lower energies, and it is often attributed to blackbody emission of the disc. In order to model the thermal Comptonization, we use the {\tt nthComp} model \citep*{Zdziarskietal_1996_nth} in {\tt XSPEC} (v.\ 12.8.2, \citealt{arnaud96}), and use disc blackbody as seed photons. 

To model the reflection component, we use {\tt relxill} v0.2h \citep{garcia14a}. That model combines the {\tt xillver} model \citep{gk10}, which uses the atomic data of {\tt XSTAR} \citep{xstar} to calculate angle-dependent reflection spectra, with the relativistic blurring code {\tt relline} \citep{dauser10}. The {\tt relxill} model allows us to calculate reflection from a disc that is truncated at an inner radius, $r_{\rm in}$. The amount of reflection is given by the relative reflection strength, ${\cal R}$, which is defined in {\tt relxill} v0.2h phenomenologically as the ratio of the flux of reflected photons to those reaching the observer directly in the 20--40 keV range (regardless of the value of $i$, see Appendix \ref{reflection}). The radial dependence of the flux incident on the disc is a power law, $\propto r^{-q}$, and we set $q=3$ unless specified otherwise. We allow the Fe abundance relative to the solar one, $A_{\rm Fe}$, to be free. 

The e-folding energy of the reflected spectrum is assumed to be $E_{\rm c}=150$\,keV while we adopt the temperature of the thermal Comptonization component of $kT_{\rm e}=100$ keV (cf.\ \citealt{wardzinski02}). The adopted value of $kT_{\rm e}$ is roughly compatible with that of $E_{\rm c}$, see \citet{z03}. Still, the value of $E_{\rm c}$ has a rather small effect because of the detector energy range of $\leq 10$ keV as well as we find the reflection strength to be at most moderate.  

In {\tt relxill}, we set the outer disc radius to $r_{\rm out}=10^3$. While the actual outer disc radius is much higher in low-mass X-ray binaries, close to the Roche lobe radius, $r_{\rm out}$ here corresponds to the outer radius of the irradiating corona, which is, most likely, much lower than that of the disc rim. Since {\tt relxill} assumes reflection from a slab, the contribution of large radii to reflection in this model is very small, and the exact value of $r_{\rm out}$ is of little importance.

We use the {\tt diskbb} model \citep{Mitsudaetal_1984}, which neglects the zero-torque inner boundary condition at the ISCO of a geometrically-thin accretion disc around a BH \citep{ss73}. Its normalization includes both the apparent disc inner radius under the above assumptions and the disc inclination. We have modified this model in order to have the true truncation radius and the inclination as input parameters, which then allows us to link them with those of the {\tt relxill} model, see Section~\ref{diskbb}. 

We use {\tt tbnew\_feo} \footnote{\url{http://pulsar.sternwarte.uni-erlangen.de/wilms/research/tbabs/}}  \citep*{Wilmsetal_2000} to model the line-of-sight absorption. We set the abundances as {\tt wilm} \citep{Wilmsetal_2000}, except that we allow the relative O abundance, $A_{\rm O}$, to be free. We set the cross sections as {\tt vern} \citep{Vernetal_1996}.

We note that some of the parameters of different models we use are physically the same.  These include the colour disc temperature of {\tt diskbb} and {\tt nthcomp}, and the photon index and the normalization of {\tt nthcomp} and {\tt relxill}. Hence, we link these parameters between the corresponding models, see Section \ref{diskbb}. In addition, some parameters are assumed to be the same for different observations, which are the disc inclination, $i$, the abundances, the absorbing column density. 

\subsection{Modification of the disc blackbody model} 
\label{diskbb}

As described in Section \ref{model} above, the spectra generated by {\tt diskbb} and {\tt relxill} depend on both the disc truncation radius, $r_{\rm in}$, and the inclination,  $i$. However, whereas {\tt relxill} depends explicitly on $r_{\rm in}$ and $i$, {\tt diskbb} depends on them through its normalization, defined as 
\begin{equation}
 N_{\tt diskbb} =\frac{(R'_{\rm in}/{1\,{\rm km}})^2 \cos i}{(D/{\rm 10\,kpc})^2},
 \label{N1}
\end{equation}
where $R'_{\rm in}$ is an apparent inner radius, defined in the case of the disc emitting un-diluted blackbody and without taking into account the zero-torque inner boundary condition. We relate $R'_{\rm in}$ to the true inner disc radius, $r_{\rm in}=\kappa^2 \zeta R'_{\rm in}/R_{\rm g}$, where $\kappa$ is the diluted-blackbody colour correction factor and $\zeta$ is the correction factor due to the inner boundary condition. Since $r_{\rm in}$ and $i$ in {\tt diskbb} and {\tt relxill} should be physically the same, we have modified the {\tt diskbb} model in {\tt XSPEC} to have its normalization defined through $r_{\rm in}$ and $i$, which  then allows linking them to the corresponding parameters of {\tt relxill}.

Furthermore, the relationship between $r_{\rm in}$ and the physical inner radius depends on the mass of the BH. This mass depends on the value of the mass function, $f$, as well as on the binary inclination, $i'$, $f={M \sin^3 i'}/{(1+M_2/M)^2}$. Given that $M_2/M \ll 1$ \citep{hynes03}, we have $f \simeq M \sin^3 i'$. If the plane of the reflecting disc is aligned with the binary plane, $i'=i$. However, $i'\neq i$ is possible if the BH is rotating around an axis different from the binary axis. In that case, we consider the BH mass, $M$, to be unrelated to $i$. In these two cases, we can express $N_{\tt diskbb}$  as a product of two main factors, one dependent on the mass, distance and the radius correction, and the other dependent on $r_{\rm in}$ and $i$. Namely,
\begin{eqnarray}
\lefteqn{
N_{\tt diskbb}\simeq 218.17 k^2 \frac{r_{\rm in}^2\cos i }{\sin^6 i} ,\quad k\equiv {f/\msun\over \kappa^2 \zeta D/1\,{\rm kpc}},
 \label{N2}}\\
\lefteqn{
 N_{\tt diskbb}= 218.17 k'^2 r_{\rm in}^2 \cos i,\quad k'\equiv {M/\msun\over \kappa^2 \zeta D/1\,{\rm kpc}}.
 \label{N3}}
\end{eqnarray}

The colour correction factor was found by \citet{st95} to be in the range of $\kappa\simeq$1.7--2, with somewhat lower values found by \citet{davis05}, see their table 1. We note that the inner disc temperature of {\tt diskbb}, $T_{\rm in}$, has now the interpretation of the colour temperature, while the effective temperature equals $T_{\rm in}/\kappa$. 

\citet{kubota98} found\footnote{\citet{gierlinski99}, using a somewhat more accurate treatment, obtained $\zeta\simeq 1/2.73\simeq 0.37$.} $\zeta\simeq 0.41$ under the assumption of the zero-torque boundary condition applying at $r_{\rm in}$. However, while it is rather likely that there is no torque at the inner radius if the disc extends to the ISCO \citep{paczynski00,shafee08}, the situation is different if the disc is truncated at $r_{\rm in}\gg r_{\rm ISCO}$. The matter in the geometrically-thin disc becomes very hot at the truncation radius \citep{liu99,rc00}, and forms an inner hot flow \citep{ADAF}, which phenomenon is unlikely to be associated with a disappearance of the torque and its related dissipation. The zero-torque condition still takes place, but only close to the horizon. Therefore, it is appropriate to include that correction factor, $\zeta$, in the soft state, where $r_{\rm in}=r_{\rm ISCO}$ is very likely. However, that factor should not be automatically included in the hard state, where the truncation radius may be $r_{\rm in}\gg r_{\rm ISCO}$.

\subsection{The considered cases}
\label{cases}

\begin{figure*}\centering{
\includegraphics[width=4.2in]{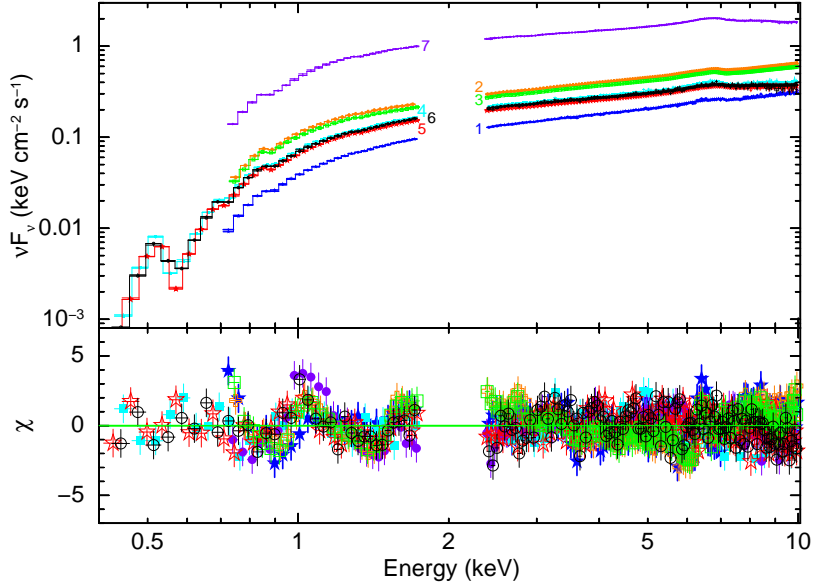} }
\caption{Results of the joint spectral fitting for all the observations using the model 1(i), see Table~\ref{tfree} for the parameters. Different colours are used for different observations, and the residuals, $\chi$, are shown in the lower panel. The symbols used are: 1: blue filled stars, 2: orange open plus signs, 3: green open squares, 4: light blue filled squares, 5: red open stars, 6: black open circles, 7: violet filled circles.
}
\label{joint}
\end{figure*}

As we have discussed above, there are a number of uncertainties in the modelling. Thus, we explore here the consequences of different assumptions. Also, the detector response at lower energies of the timing mode data is not well known. Even when that response is known, the soft X-ray energy range is physically complex, e.g., due to the effect of the irradiation of disc by the hot phase \citep*{gierlinski08,gierlinski09}. Hence, we also consider fitting results where the soft band is not taken into account. We list our various sets of the assumptions below\footnote{Some further cases are discussed in \url{http://arxiv.org/abs/1512.01833v1}.}. We assume $D=8$ kpc. 
\begin{itemize}
\item Case 1. We fit the full energy bands, 0.4--10\,keV in the imaging mode and 0.7--10\,keV in the timing mode (neglecting in both cases the 1.75--2.35\,keV range).
\begin{enumerate}
 \item We use equation (\ref{N2}) for the {\tt diskbb} normalization with a free value of $k$, but assumed to be the same for all observations. The obtained value of $k$ can be then interpreted as corresponding to various combinations of $f$, $D$ and $\kappa^2 \zeta$. 
\item We fix $f=5.8\msun$, $\kappa=1.7$ and $\zeta=0.41$ in equation (\ref{N2}), and then $M=10\msun$, $\kappa=1.7$, $\zeta=1$ in equation (\ref{N3}).
\item We use the standard {\tt diskbb} normalization and allow it to be free and different for each data set. 
\end{enumerate}
\item Case 2. We fit in the reduced energy band of $E>2.35$ keV and fix the parameters for the absorption and disc components at those found for the full energy band. We further consider various assumptions.
\begin{enumerate}
 \item The condition for tying the $r_{\rm in}$ and $i$ of {\tt diskbb} and {\tt relxill} is lifted. The {\tt diskbb} model is used without any modification, and we fit {\tt relxill} for $r_{\rm in}$ and $i$ independently. The parameters of the {\tt diskbb} are frozen to those obtained in case 1(i). We also study a case with free individual parameters of the {\tt diskbb}, which is analogous to case 1(iii).
 \item We add an additional un-blurred reflection component. This is motivated by the possibility of reflection from a distant medium in addition to that corresponding to a flat disc, e.g., taking into account a disc flaring. 
 \item We assume the lamppost geometry \citep{mm96,mf04,dauser14} using the {\tt relxilllp} model.
\end{enumerate}
\end{itemize}

\section{Results}
\label{results}

\begin{figure}\centering
{\includegraphics[width=\columnwidth]{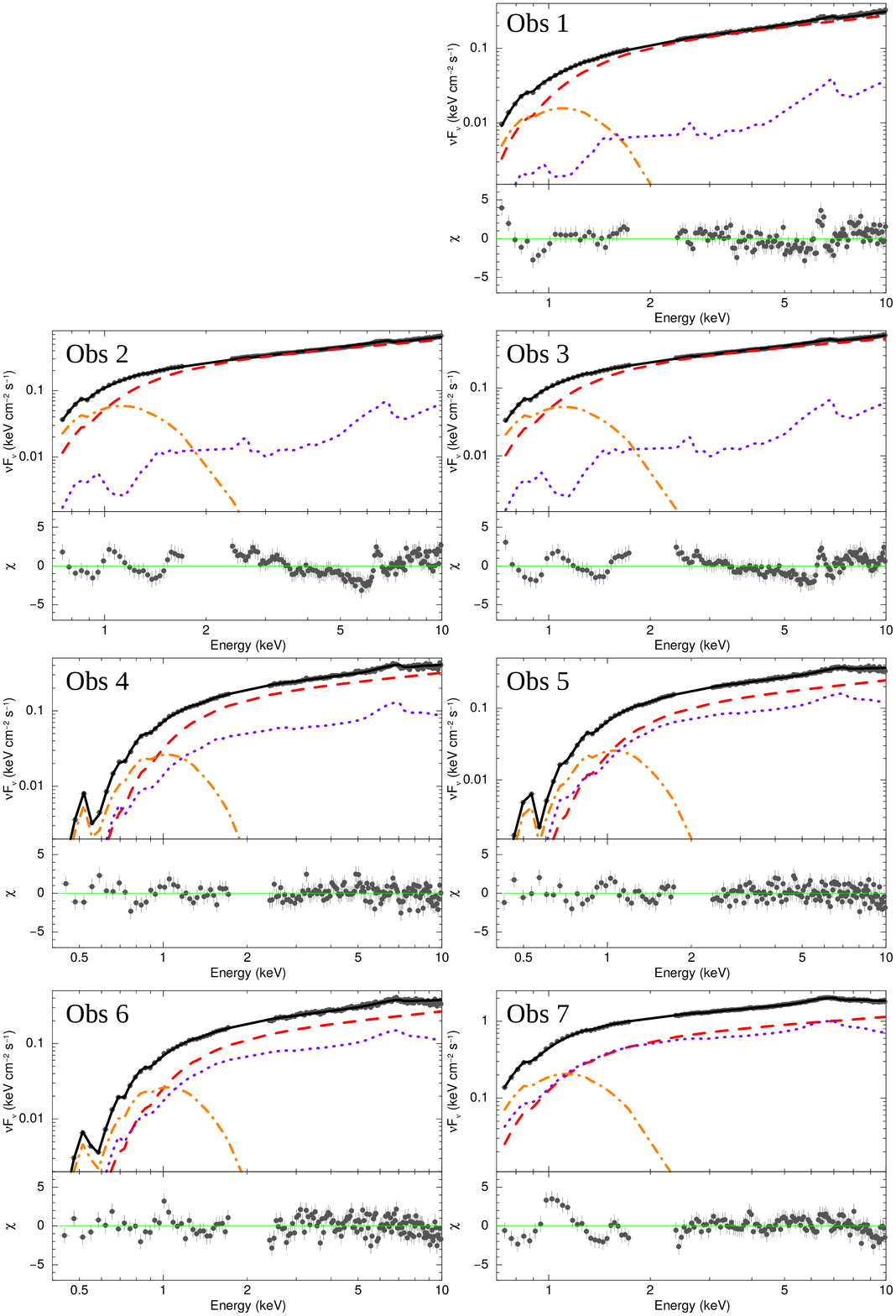} }
\caption{Results of the joint spectral fitting for individual observations for the case 1(i), see Table~\ref{tfree}. The residuals, $\chi$, of the fits are shown in the lower panels. The orange dot-dashed, red dashed, violet dotted and black solid curves show the disc blackbody, incident continuum, reflection and the total model, respectively.
}
\label{ind}
\end{figure}

\begin{table*}\centering
\caption{The fitted model parameters of the seven observations of GX 339--4 for the case 1(i). The model is {\tt tbnew\_feo(diskbb+nthcomp+relxill)}. The {\tt diskbb} model is modified to incorporate $r_{\rm in}$ and $i$ as input parameters, see equation (\ref{N2}), which are then linked with the corresponding parameters of {\tt relxill}. The factor $k$, equation (\ref{N2}), is allowed to be free ($\simeq 0.14$); the value of $f$ given below assumes $\kappa=2$, $\zeta=1$, $D=8$ kpc. }
\begin{tabular}{cccccccc}
\hline
Parameter       & Obs. 1     & Obs. 2     & Obs. 3     & Obs. 4     & Obs. 5     & Obs. 6     & Obs. 7     \\
\hline
$N_{\rm H}\,(10^{22}\,{\rm cm}^{-2})$& \multicolumn{7}{c}{$0.70\pm0.01$}\\
$A_{\rm O}$       & \multicolumn{7}{c}{$1.50\pm0.03$}\\
$f\,(\msun)$      & \multicolumn{7}{c}{$4.5_{-0.7}^{+1.1}$}\\
$kT_{\rm in}$ (keV)     & $0.19\pm0.004$   &$0.20\pm0.003$   & $0.20\pm0.002$   &$0.17\pm0.004$   &$0.18\pm0.004$   & $0.17\pm0.004$   &$0.21\pm0.003$   \\
$\Gamma$        & $1.54\pm0.01$   &$1.60\pm0.01$   & $1.60\pm0.01$   &$1.64\pm0.01$   &$1.60\pm0.01$   & $1.62\pm0.02$   &$1.69\pm0.01$    \\
$N_{\tt nthcomp}$     & $0.09\pm0.002$   &$0.22\pm0.002$   & $0.20\pm0.002$   &$0.14\pm0.01$   &$0.10\pm0.01$   & $0.11\pm0.02$   &$0.52\pm0.01$    \\
$i\,(\degr)$       & \multicolumn{7}{c}{$44.6_{-5.4}^{+4.7}$}\\
$r_{\rm in }$ ($R_{\rm g}$)   & $23.3_{-7.1}^{+11.8}$&$38.0_{-13.6}^{+19.2}$& $40.4_{-12.7}^{+20.5}$ &$52.2_{-18.1}^{+25.6}$&$42.5_{-15.2}^{+20.2}$& $47.5_{-16.8}^{+23.1}$&$68.5_{-24.9}^{+23.0}$ \\
$\log_{10}\xi$ (erg\,cm\,s$^{-1}$)  & $2.34_{-0.08}^{+0.11}$ &$2.31\pm0.04$   & $2.32_{-0.04}^{+0.05}$ &$3.03_{-0.10}^{+0.07}$ &$3.18_{-0.08}^{+0.10}$ & $3.20_{-0.11}^{+0.18}$ &$3.30_{-0.03}^{+0.01}$ \\
$A_{\rm Fe}$       & \multicolumn{7}{c}{$0.83_{-0.04}^{+0.05}$}\\
${\cal R}$       & $0.38\pm0.06$   &$0.32_{-0.04}^{+0.03}$ & $0.33\pm0.03$    &$0.50_{-0.11}^{+0.15}$ &$0.75_{-0.16}^{+0.20}$ & $0.67_{-0.18}^{+0.39}$ &$>0.94$     \\
$\chi^2/\nu$ & \multicolumn{7}{c}{1371.7/1046}\\
\hline
\label{tfree}
\end{tabular}
\end{table*}

\begin{figure}\centering
{\includegraphics[width=\columnwidth]{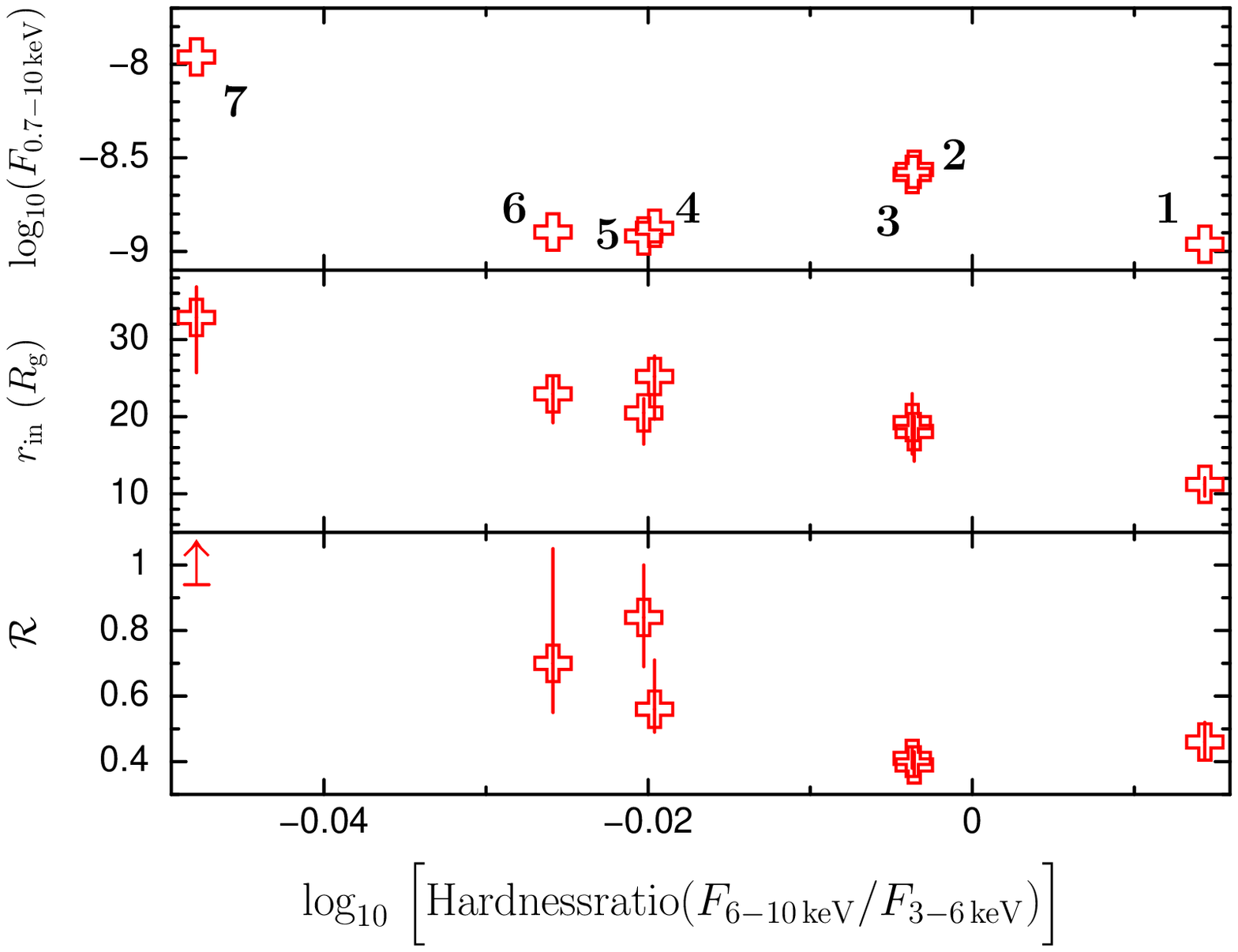} }
\caption{The hardness-flux diagram (top) of all the observations in the case 1(i), see Table~\ref{tfree}. The corresponding values of $r_{\rm in}$ (middle) and ${\cal R}$ (bottom) vs.\ the hardness ratio.}
\label{H_F}
\end{figure}

\subsection{Spectral study in the full energy bands}
\label{full}

We allow here the normalization of the {\tt diskbb} to be free, but we separate it into two factors, one related to either $f$ or $M$, $D$ and $\kappa^2\zeta$, and one related to $r_{\rm in}$ and $i$, see equations (\ref{N2}), (\ref{N3}). The latter is independently determined by {\tt relxill}. We then impose a single value of either $k$ or $k'$ for all the data. 

We first assume a free $k$, case 1(i), see Table \ref{tfree} and Figs.\ \ref{joint}, \ref{ind}. We find expected values for the normalization factor, $k$, and the inclination, $i$. The value of is $k\simeq 0.14^{+0.05}_{-0.03}$,  which is only about 20 per cent away at the best fit and compatible within $1\sigma$ with  $k\simeq 0.18$, which corresponds to our default values of $f=5.8\msun$, $\kappa=2$, $\zeta=1$ and  $D=8$ kpc. The fitted inclination of $i\simeq 45\degr\pm 5\degr$ corresponds to $M\simeq (12$--25)$\msun$ at $f=5.8\pm 0.5\msun$. We also obtain the Fe abundance of the reflector close to the solar. The truncation radii are $r_{\rm in}\gg r_{\rm ISCO}$ for all of the observations, which justifies the choice of $\zeta=1$ (see Section \ref{diskbb}), and is compatible with the truncated  disc geometry in the hard state. The fractional reflection strength is $\lesssim 1$ for all the data. While it is compatible with that geometry, it implies some overlap of the hot flow with the disc, or the presence of some cold clumps within the hot flow. The fitted $N_{\rm H}$ is comparable to that obtained by other investigations, e.g., KDD14, \citet{plant14b}. Also, the inner disc temperature is $kT_{\rm in}\simeq 0.2$ keV for all the observations, which implies that the colour correction, $\kappa$, should be correspondingly approximately constant. The spectral index is $\Gamma\simeq 1.5$--1.7, which is typical for the hard state. The total reduced $\chi^2$ is also acceptable.

Then, we assume the values of $\zeta=0.41$ \citep{kubota98} and $\kappa=1.7$ \citep{st95}, and the default values of $f$ and $D$, case 1(ii). We find much smaller values of $r_{\rm in}$, of $\simeq 3$--$10$, much closer to $r_{\rm ISCO}$ than before, but the fit strongly worsens, to $\chi^2=1810/1047$, i.e., $\Delta\chi^2=+438$ for adding one d.o.f. Thus, we rule out this model as statistically unacceptable. As another variant, we consider the inclination of the inner disc being different from that of the binary orbit, $i'\neq i$. We use equation (\ref{N3}) with $M=10\,\msun$, $\kappa=1.7$, $\zeta=1$.  This model yields the best fit parameters and the $\chi^2$ almost identical to those of case 1(i), Table~\ref{tfree}. It differs from case 1(i) simply by a different interpretation of the normalization of {\tt diskbb}. 

Fig.\ \ref{joint} shows the unfolded spectra, the models and the residuals for case 1(i) for all of the observations together. We use the $\nu F_\nu=E F_E$ representation, with $F$ given in units of keV. The individual fits with their residuals are shown in Fig.\ \ref{ind}. We can see that the residuals of the data taken in imaging mode (observations 4--6) are acceptable. However, those of the timing mode, especially those of observations 2 and 3, show relative large structure. Specifically, the excess at 1\,keV is prominent. A similar structure has been reported by KDD14, who argued that this probably indicates that the soft spectrum is not well modelled by {\tt diskbb}.

We then consider the hardness-flux diagram for the model 1(i). We calculate the unabsorbed energy flux in 0.7--10\,keV energy band, and the hardness ratio as the energy flux ratio between the 6--10\,keV band to that in the 3--6\,keV band, see the top panel of Fig.\ \ref{H_F}. We find the flux to be only approximately anti-correlated with the hardness ratio, see Section \ref{hysteresis}. An overall anti-correlation is expected given that the flux increase in the hard state leads eventually to a transition to the soft state. The middle and bottom panels of Fig.\ \ref{H_F} show $r_{\rm in}$ and ${\cal R}$, respectively, as functions of the hardness ratio. We find the reflection strength is decreasing as the spectrum becomes harder, as expected in the truncated disc model. 

However, we unexpectedly find $r_{\rm in}$ to anti-correlate with the hardness. This is contrary to the truncated disc model, in which $r_{\rm in}$ decreases during the hard state of the outburst, as the spectrum becomes softer and the luminosity increases. It also disagrees with models in which $r_{\rm in}=r_{\rm ISCO}$. We consider this anti-correlation to be an artefact caused by the adopted assumptions of our {\tt diskbb} modelling. In our modified model, we impose the factor of the disc normalization related to the mass, the distance, the inclination and the colour correction to be the same for all the observations. Physically, the increase of the flux in the disc component should occur by an increase of $T_{\rm in}$, both  because of the increase of the accretion rate and the expected decrease of $r_{\rm in}$. However, we obtain $T_{\rm in}$ to be almost constant for all the seven observations, in spite of the flux  increase by an order of magnitude. Then, the only way to increase the disc blackbody flux is to increase the inner radius. This demonstrates that while our model accounts well for the overall properties of the hard state, giving likely values of the inclination and the Fe abundance, it does not account for the spectral variability within that state. The apparent reason for this is that the {\tt diskbb} model is not the proper description of the soft X-ray component present in the data. We discuss possible causes of it in Section \ref{caveats}.

Given the apparently unphysical relationship between $r_{\rm in}$ and the hardness ratio, we also consider the case 1(iii), with the {\tt diskbb} normalization allowed to vary between the observations. This corresponds to treating this spectral component as phenomenological, without imposing any physical constraints. We obtain a significant improvement of the fit, $\Delta\chi^2=-222$ for 6 less d.o.f. However, the best-fit Fe abundance is unphysical, namely $A_{\rm Fe}\simeq 20$; also  $i\simeq 29\degr$. The reflection is, however, very weak, ${\cal R}\lesssim 0.1$, while the values of $r_{\rm in}$ span the range of about 6--110, still $\gg r_{\rm ISCO}$ for our assumed $a=0.9$. Even at that value of $A_{\rm Fe}$, the model still implies a truncated disc geometry. We have also tried a similar fitting with the $A_{\rm Fe}$ now fixed to 1.58, the value found for the best fit case by \citet{furst15}. Then we obtain $\chi^2$/dof = 1293/1041, with $i=29\degr$--$35\degr$. The disc is still highly truncated with $r_{\rm in}$ in the range 26--215, and we obtain low values of the reflection fraction.

\begin{table*}\centering
\caption{The model parameters for the case 2(i), in which the values of $r_{\rm in}$ fitted by {\tt relxill} are allowed to be different from those implied by the normalization of {\tt diskbb}.}
\begin{tabular}{cccccccc}
\hline
Parameter       & Obs. 1     & Obs. 2     & Obs. 3      & Obs. 4     & Obs. 5     & Obs. 6     & Obs. 7     \\
\hline                                               
$N_{\rm H}\,(10^{22}\,{\rm cm}^{-2})$&\multicolumn{7}{c}{$0.71^{(f)}$}\\ 
$A_{\rm O}$       &\multicolumn{7}{c}{$1.48^{(f)}$}\\   
$kT_{\rm in}$ (keV)     & $0.19^{(f)}$   & $0.20^{(f)}$   & $0.19^{(f)}$    & $0.16^{(f)}$   & $0.17^{(f)}$   & $0.17^{(f)}$    & $0.20^{(f)}$    \\
$\Gamma$        & $1.49\pm 0.01$   & $1.56\pm 0.004$   & $1.56\pm 0.004$   & $1.62\pm 0.01$   & $1.58\pm 0.01$   & $1.61\pm 0.01$   & $1.69\pm 0.005$   \\
$N_{\tt nthcomp}$     & $0.09\pm 0.001$   & $0.21\pm 0.002$   & $0.19\pm 0.002$   & $0.12\pm 0.01$   & $0.08\pm 0.01$   & $0.08\pm 0.02$   & $0.52\pm 0.06$   \\
$i\,(\degr )$      &\multicolumn{7}{c}{$43.8_{-5.4}^{+6.1}$}\\
$r_{\rm in}\,(R_{\rm g}$)   &$600_{-325}^{+\infty}$     &$226_{-94.2}^{+195}$ &$166_{-77.0}^{+114}$ & $275_{-196}^{+\infty}$  & $50.3_{-32.1}^{+\infty}$    & $286_{-233}^{+\infty}$     & $42.0_{-14.1}^{+18.6}$ \\
$\log_{10} \xi$\,(erg\,cm\,s$^{-1}$)  &$2.40_{-0.16}^{+0.32}$ & $2.70_{-0.16}^{+0.03}$ & $2.69_{-0.15}^{+0.05}$  & $3.19_{-0.09}^{+0.07}$ & $3.31_{-0.09}^{+0.06}$ & $3.34\pm0.03$   & $3.31_{-0.03}^{+0.02}$ \\
$A_{\rm Fe}$       &\multicolumn{7}{c}{$1.00_{-0.06}^{+0.15}$}\\ 
${\cal R}$       & $0.16\pm 0.04$   & $0.13\pm0.02$   & $0.14\pm0.01$    & $0.56_{-0.11}^{+0.21}$ & $0.95_{-0.21}^{+0.30}$ & $0.98_{-0.34}^{+0.42}$ & $>0.89$     \\
$\chi^2/\nu$ & \multicolumn{7}{c}{771/843}\\
\hline
\end{tabular}
\label{t_best}
\end{table*}
 
\subsection{Spectral fits in the reduced energy band}
\label{2.35}

As we discussed above, the detector calibration at soft X-rays is less certain than that at hard X-rays, especially for the timing mode. Also, the modelling at soft X-rays is complex and uncertain. Therefore, we perform fitting also in a reduced energy band, ignoring the low-energy, $<2.35$ keV, part of all of the data. Given the narrowness of this band, we assume the parameters of the absorber and the {\tt diskbb} component to be fixed at the values obtained by including the low-energy band.

\begin{figure}\centering
{\includegraphics[width=\columnwidth]{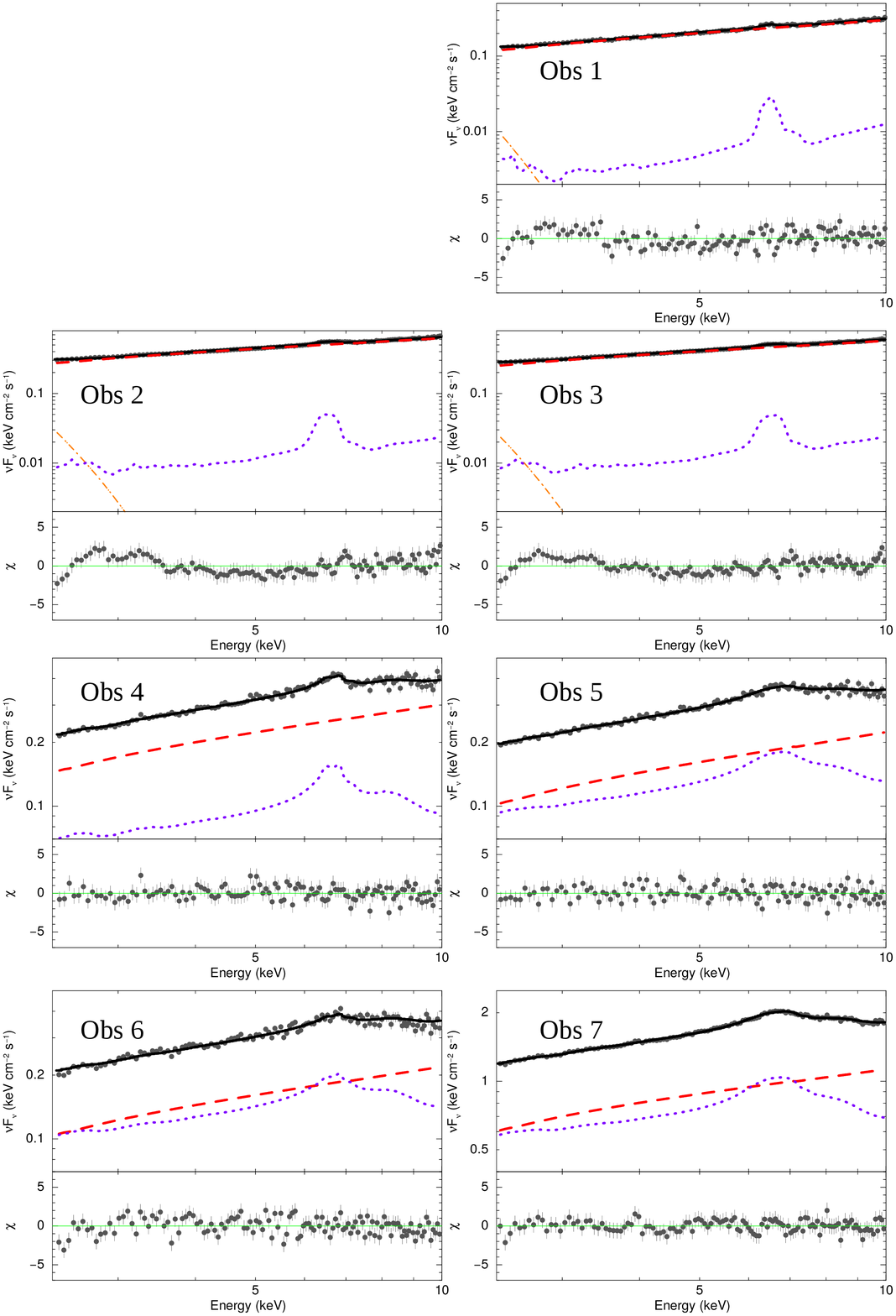} }
\caption{The spectra obtained in the case 2(i), see Table~\ref{t_best}. The orange dot-dashed, red dashed, violet dotted and black solid curves show the disc blackbody, incident continuum, reflection and the total model, respectively.
}
\label{2(i)}
\end{figure}

\begin{figure}\centering
{\includegraphics[width=7cm]{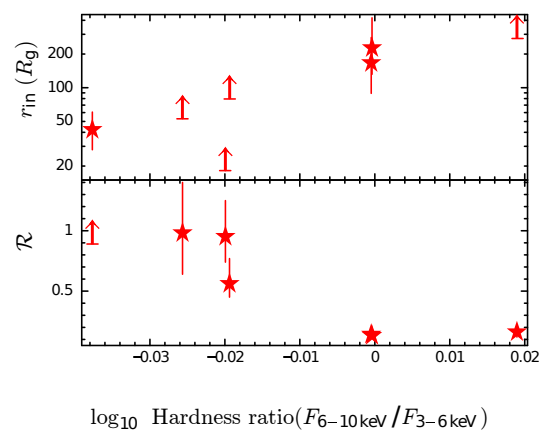} }
\caption{The values of $r_{\rm in}$ (top) and $\cal{R}$ (bottom) as functions of the hardness ratio for the case 2(i). This figure can be compared with Fig.\ \ref{H_F}. The dependencies of the 0.3--10 keV flux on the hardness, shown in Fig.\ \ref{H_F}(top), are very similar for the cases shown here.
}
\label{rin_R}
\end{figure}

Case 2(i). We relax the requirement that the disc inner radius is the same in both {\tt diskbb} and {\tt relxill}. The results are shown in  Table~\ref{t_best} and Fig.~\ref{2(i)}. We note that as the $r_{\rm in}$ of {\tt relxill} is no more tied to the {\tt diskbb}, that value is less constrained now, and only lower limits are obtained in four cases. However, the values are consistently large. The dependencies of $r_{\rm in}$ and $\cal R$ on the hardness ratio are shown in Fig.~\ref{rin_R}. We see now the expected increase of $r_{\rm in}$ with the hardness.

As a variant of this case, we have also tried a fit with $r_{\rm in} = r_{\rm ISCO}$, but obtained an unacceptable $\chi^2$/dof = 1216/850. As another variant, we have considered free individual values of the normalization of {\tt diskbb}, as in the case 1(iii). However, unlike that case (where we fitted the full energy band), we obtain not only a significantly better $\chi^2$/dof = 625/829, but also physically plausible fit parameters, namely $i=45\degr$--$58\degr$, $A_{\rm Fe}=1.0\pm 0.2$, $r_{\rm in}= 147_{-11}^{+64}$, $165_{-38}^{+54}$, $198_{-101}^{+123}$, $994_{-859}^{+\infty}$, $78_{-42}^{+460}$, $453_{-350}^{+527}$, $57_{-13}^{+18}$ for  for the obs.\ 1--7, respectively. These high values of $r_{\rm in}$ support the truncated disc scenario. We have also tried a model without the {\tt diskbb} component, but found a large $\chi^2$/dof = 932/836. Clearly, the presence of a soft component is statistically required, even at $E\geq 2.35$ keV.

\begin{table*}\centering
\caption{The model parameters for the case 2(ii), which includes an additional 
un-blurred reflection component from a distant medium, e.g., a flared disc. 
The model is {\tt tbnew\_feo(diskbb+nthcomp+xillver+relxill)}. Subscript `u' denotes the parameters of the un-blurred {\tt xillver} component.}
\begin{tabular}{cccccccc}
\hline
Parameter          & Obs. 1     & Obs. 2     & Obs. 3     & Obs. 4     & Obs. 5     & Obs. 6     & Obs. 7     \\
\hline                                               
$N_{\rm H}\,(10^{22}\,{\rm cm}^{-2})$   &\multicolumn{7}{c}{$0.71^{(f)}$}\\ 
$A_{\rm O}$         &\multicolumn{7}{c}{$1.48^{(f)}$}\\   
$kT_{\rm in}$ (keV)       & $0.19^{(f)}$   & $0.20^{(f)}$   & $0.19^{(f)}$   & $0.16^{(f)}$   & $0.17^{(f)}$   & $0.17^{(f)}$    & $0.20^{(f)}$    \\
$\Gamma$          & $1.47\pm 0.01$   & $1.55\pm 0.05$   & $1.56\pm 0.004$   & $1.61\pm 0.01$   & $1.58\pm 0.01$   & $1.61\pm 0.02$   & $1.67\pm 0.02$   \\
$N_{\tt nthcomp}$        & $0.07\pm 0.02$   & $0.20\pm 0.02$   & $0.19\pm 0.02$   & $0.10\pm 0.02$   & $0.06\pm 0.01$   & $0.06_{-0.003}^{+0.10}$ & $0.40\pm 0.12$   \\
$i\,(\degr )$         &\multicolumn{7}{c}{$45.2_{-3.7}^{+4.9}$}\\
$r_{\rm in}\,(R_{\rm g}$)      & $>293$    & $227_{-84.0}^{+211}$ & $144_{-96.4}^{+107}$ & $>94.0$     & $<7.1$     & $>132$     & $19.5_{-8.0}^{+15.0}$\\
$\log_{10} \xi$\,(erg\,cm\,s$^{-1}$)    & $2.41_{-0.11}^{+0.30}$ & $2.69_{-0.16}^{+0.03}$ & $2.69_{-0.13}^{+0.05}$ & $3.02_{-0.47}^{+0.62}$ & $3.29_{-0.14}^{+0.08}$ & $3.28_{-0.09}^{+0.04}$ & $3.33_{-0.02}^{+0.04}$ \\
${\cal R}$         & $0.19\pm0.04$   & $0.13\pm0.02$   & $0.13_{-0.04}^{+0.03}$ & $0.22_{-0.09}^{+0.32}$ & $>0.90$     & $>0.84$     & $>0.96$     \\
$\log_{10} \xi_{\rm u}$\,(erg\,cm\,s$^{-1}$)  & $>3.69$     & $4.7$     & $2.02\pm0.39$   & $3.36_{-0.07}^{+0.08}$ & $3.24_{-0.10}^{+0.09}$ & $>3.71$     & $3.34_{-0.20}^{+0.10}$ \\
${\cal R}_{\rm u}$       & $0.16_{-0.02}^{+0.17}$ & $<0.04$     & $<0.05$     & $0.56_{-0.19}^{+0.17}$ & $0.43_{-0.11}^{+0.24}$ & $0.48_{-0.14}^{+0.29}$ & $0.54_{-0.12}^{+0.09}$     \\
$A_{\rm Fe}$         &\multicolumn{7}{c}{$0.95_{-0.06}^{+0.07}$}\\
$\chi^2/\nu$         & \multicolumn{7}{c}{752/829}\\
\hline
\end{tabular}
\label{t_unblurred}
\end{table*}

\begin{table*}\centering
\caption{The model parameters for the case 2(iii), with the lamppost geometry. The model is {\tt tbnew\_feo(diskbb+relxilllp)}. Given the weak fitting constraints, we set $r_{\rm in}=r_{\rm ISCO}$.}
\begin{tabular}{cccccccc}
\hline
Parameter                            & Obs. 1                  & Obs. 2                  & Obs. 3                     & Obs. 4                  & Obs. 5                  & Obs. 6                   & Obs. 7                   \\
\hline                                                                                                                                                                                            
$N_{\rm H}\,(10^{22}\,{\rm cm}^{-2})$&\multicolumn{7}{c}{$0.71^{(f)}$}\\  
$A_{\rm O}$                          &\multicolumn{7}{c}{$1.48^{(f)}$}\\          
$kT_{\rm in}$ (keV)                  & $0.19^{(f)}$            & $0.20^{(f)}$            & $0.19^{(f)}$               & $0.16^{(f)}$            & $0.17^{(f)}$            & $0.17^{(f)}$             & $0.20^{(f)}$             \\
$\Gamma$                             & $1.44\pm 0.01$          & $1.49\pm 0.005$         & $1.50\pm 0.004$            & $1.58\pm 0.01$          & $1.56\pm 0.005$         & $1.59\pm 0.01$           & $1.67\pm 0.005$          \\
$N_{\tt relxilllp}$                  & $0.08\pm 0.001$         & $0.20\pm 0.003$         & $0.19\pm 0.003$            & $0.11\pm 0.01$          & $0.08\pm 0.02$          & $0.09\pm 0.02$           & $0.54\pm 0.04$           \\
$i\,(\degr )$                        &\multicolumn{7}{c}{$40.25_{-2.79}^{+3.36}$}\\
$h\,(R_{\rm g})$                     & $>900$                  & $366_{-181}^{+605}$     & $188_{-76}^{+315}$         & $>119$                  & $76_{-43}^{+403}$       & $>2$                      & $52_{-11}^{+22}$      \\
$\log_{10} \xi$\,(erg\,cm\,s$^{-1}$)      & $2.32_{-0.04}^{+0.06}$  & $2.67_{-0.15}^{+0.01}$  & $2.63_{-0.12}^{+0.08}$     & $3.16\pm0.03$           & $3.28_{-0.04}^{+0.02}$  & $3.29_{-0.03}^{+0.02}$    & $3.27_{-0.03}^{+0.04}$   \\
$A_{\rm Fe}$                         &\multicolumn{7}{c}{$0.84\pm 0.03$}\\  
${\cal R}$                           & $0.23_{-0.03}^{+0.01}$  & $0.14_{-0.01}^{+0.02}$  & $0.17_{-0.02}^{+0.01}$     & $0.61_{-0.05}^{+0.03}$  & $0.91_{-0.05}^{+0.01}$  & $0.85_{-0.08}^{+0.04}$    & $>0.97$                  \\
$\chi^2/\nu$ & \multicolumn{7}{c}{883/843}\\
\hline
\end{tabular}
\label{t_lamppost}
\end{table*}

Case 2(ii). We then test for the presence of an unblurred reflection component. This is expected, given that the standard accretion disc of \citet{ss73} is flared, with the height to radius ratio of $H/R\propto R^{1/8}$, and with a larger exponent in the presence of irradiation. This effect is not included in {\tt relxill}, which assumes a flat reflector. We can take it into account by including another reflection component prior to relativistic blurring, as given by the {\tt xillver} model \citep{gk10}. The results are shown in Table~\ref{t_unblurred}. We see the fits are relatively poorly constrained, which is due to the absence of strong relativistic effects in the data, as we found before. Thus, adding an unblurred reflection to a weakly blurred one does not allow a determination of the parameters of both. Still, we find high truncation radii and weak reflection in all the data. For obs.\ 5, we obtain an upper limit and the disc can in principle extend down to the ISCO, but the data are also consistent with a truncated disc. Statistically, this additional component does not improve the fit, we obtain $\Delta\chi^2 \simeq -19$ by adding 14 more free parameters to the model 2(i). For this case, we have also tried a fit with $r_{\rm in}= r_{\rm ISCO}$. The $\chi^2$/dof = 795.5/836, i.e., $\Delta\chi^2 \simeq +43$ with 7 more dof. In all cases the reflection fraction is low for both {\tt xillver} (0.06--0.73) and {\tt relxill} (0.10--0.89).

Case 2(iii). We now consider the lamppost geometry \citep{mm96,mf04,dauser14}, in which a power-law point source is located on the BH rotation axis and perpendicular to a flat disc surrounding the BH. It is implemented in the {\tt relxilllp} model. The model has been relatively widely used, e.g., by \citet{furst15} for GX 339--4. The radial profile of the flux irradiating the disc is calculated using GR given the value of the height of the source, $h$. Similarly to the case 2(ii), we have found that this model is weakly constrained by the data, with very large ranges of the allowed values of $r_{\rm in}$. Therefore, we have fixed $r_{\rm in}=r_{\rm ISCO}$, which model yields a $\Delta\chi^2$ of only $+3$ with respect to free $r_{\rm in}$ case. The results are given in Table~\ref{t_lamppost}. Thus, the model is consistent with the disc extending down to the ISCO, but also a truncated disc is allowed. The latter case may be indicated by the relatively low values of the reflection fraction, in particular for the observations 1--3. However, when the disc extends to the ISCO, the height of the source is large in most cases, $h\gg 1$, which implies that relativistic effects are still weak, as the innermost parts of the disc are only weakly irradiated. The statistical quality of the fit of this model is poor, $\chi^2$ is high and much worse, $\Delta\chi^2=+112$, than that of the case 2(i). Clearly, this model is not statistically preferred. We have also added an un-blurred component, i.e., due to reflection from a distant medium subtending a substantial solid angle, to this model. This has not improved the statistics, yielding $\chi^2/{\rm d.o.f.}\simeq 871/829$, i.e., an increased reduced $\chi^2$ compared to the case without the un-blurred reflection. 

A further possible complexity possible is to allow for a broken power-law radial profile of the flux incident on the disc. Thus, we allow the emissivity to be $\propto r^{-q_1}$ for $r \leq r_{\rm br}$ and $\propto r^{-q_2}$ for $r \geq r_{\rm br}$. We have found that the presence of the additional parameters causes the fit to be rather weakly constrained. Thus, we have fixed the outer index at its standard value, $q_2=3$. We have then found that the best-fit values of $q_1$ are also close to 3, and, consequently, the model gives results similar to our default case, 2(i). In particular, the disc inner radii are still large, ranging from 42 to 600 at the best fit, and $i\simeq 44\degr\pm 4\degr$, $A_{\rm Fe}\simeq 1.0\pm 0.1$, with the $\chi^2/{\rm d.o.f.}\simeq 771/836$. The reduced $\chi^2$ is larger than in the case 2(i), which shows that this model is not statistically preferred. Finally, we have also explored the effect of ionization dependent on the radius, which complexity is allowed in the {\tt relxill}. This again has not improved the fit and did not lead to constrained results. Similarly to most of the cases above, the obtained values of $r_{\rm in}$ were high.

Our results of large truncation radii are consistent with those of \citet{demarco15}, which were obtained based on X-ray reverberation time lags, in which the X-ray source irradiates the disc. Those authors analysed the \xmm\/ observations 1, 2--3 (combined) and 7, and obtained the time lags decreasing from $\simeq$14 to $\simeq$4 ms as the bolometric Eddington ratio increases from 0.027 to 0.152 (see Table \ref{tobs}). In the truncated disc model, these lags correspond to $r_{\rm in}$ decreasing from $\simeq (280\pm 70) (M/10\msun)^{-1}$ to $\simeq (60\pm 25) (M/10\msun)^{-1}$, which fully agree with the results for our best model, case 2(i) (Table \ref{t_best}).

\section{Discussion}
\label{discussion}

\begin{figure}\centering{
\includegraphics[width=7.cm]{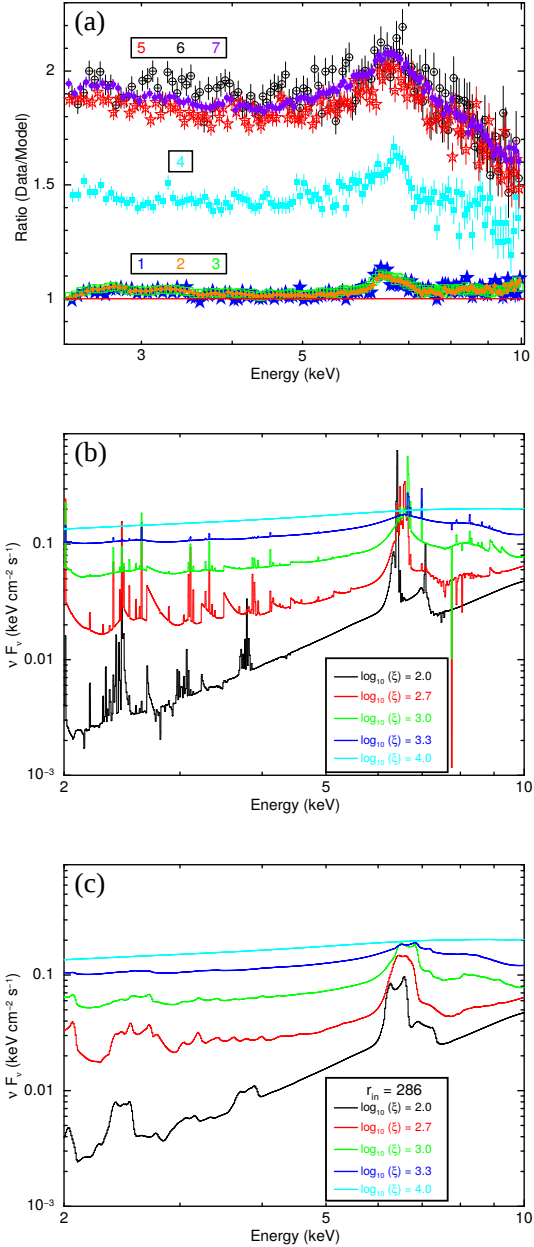} }
\caption{(a) The data-to-model ratios for the model without the reflection component and for case 2(i). The symbols (the same as in Fig.~\ref{joint}) are: 1: blue filled stars, 2: orange open plus signs, 3: green open squares, 4: light blue filled squares, 5: red open stars, 6: black open circles, 7: violet filled circles. (b) The reflection model without relativistic blurring ({\tt xillver}) for selected values of the ionization parameter: $\log_{10} \xi=2$, 2.7, 3, 3.3 and 4, with the other parameters at the best fit values of obs.\ 6 (see Table~\ref{t_best}). (c) The same as in panel (b) but with relativistic broadening for $r_{\rm in}=286$.
}
\label{ionization_broadening}
\end{figure}

\subsection{Effect of ionization on line broadening}
\label{ionization}

As we have shown in Fig.~\ref{Fe_line}, the width of the Fe K line complex increases with the decreasing spectral hardness, and the line for obs.\ 5, 6, 7 appears broad. On the other hand, we always find a significantly truncated disc in our analysis. The effect we found responsible for the line broadening in addition to the relativistic effects is ionization. We illustrate it for our case 2(i). In order to show the relative contribution of the reflection spectrum (for {\tt relxill}), we omit it, and then show the resulting data-to-model ratio in Fig.~\ref{ionization_broadening}(a). We note that the obs.\ 1--3 have both narrow Fe lines and low relative reflected continua, whereas the obs.\ 5--7 have both broad lines and high reflected continua, with the obs.\ 4 being intermediate between those two cases. The level of the reflected continuum increases with the ionization parameter (as the disc is more reflective when more ionized) and with the reflection fraction. Indeed, both ${\cal R}$ and $\log_{10}\xi$ are much higher for obs.\ 5--7 than for 1--3, see Table \ref{t_best}. On the other hand, the value of $r_{\rm in}$ is high for all of the cases, though there is a decreasing trend with the softness, see Fig.\ \ref{rin_R}.

Thus, we investigate the effect of the ionization on the line broadening. We show the the rest-frame reflected spectra (i.e., without relativistic effects), given by {\tt xillver}. We show the model reflected spectra in Fig.~\ref{ionization_broadening}(b) for $\log_{10}\xi=2$, 2.7, 3, 3.3 and 4. Specifically the values of $\log_{10} \xi= 2.7$ and 3.3 correspond closely to our observations with the narrow and broad lines, see Table \ref{t_best}. We clearly see that the width of the Fe K complex increases with the increasing $\xi$. Then we show how the relativistic effects further broaden the line in Fig.~\ref{ionization_broadening}(c). We use the {\tt relxill} model for the same values of $\xi$ and for $r_{\rm in}=286$, which is the highest value among the broad-line observations, 5--7, corresponding to the obs.\ 6. We see some rather modest further broadening. Hence, we conclude that the disc ionization plays a vital role in broadening the line. We give below a brief qualitative explanation of the physical processes that can lead to this effect, following \citet{garcia13}.

In {\tt xillver}, the reflected spectrum is produced from a plane-parallel slab of uniform density gas irradiated from the top by a power-law spectrum with an exponential high-energy cutoff. The ionization parameter is defined as $\xi\equiv 4\upi F_{\rm X}/n_{\rm e}$, where $F_{\rm X}$ is the flux integrated over 1--1000 Ry and $n_{\rm e}$ is the electron density. The source function of the radiation transfer equation is given by the ratio of the total emissivity to the total opacity. The opacity consists of the scattering term, $\alpha_{\rm KN}(E)$ (given by $n_{\rm e}$ times the Klein-Nishina cross section), and the atomic absorption term. The emissivity consists of two parts, the continuum and line emissivity, $j(E)$, and one due to Comptonization, equal to $\alpha_{\rm KN}(E)$ times the Comptonized mean intensity, $J_{\rm c}$. The probability of Compton scattering of a photon with energy, $E$, is approximated by a Gaussian profile (cf.\ \citealt*{ross_78, ross_fabian93}) with the centre at $E_0 =E+\Delta E$ and the half-width, $\sigma$, where
\begin{equation}
 \frac{\Delta E}{E} = \frac{4kT}{m_{\rm e}c^2}-\frac{E}{m_{\rm e}c^2},\quad 
\sigma=E\left[{2kT\over m_{\rm e}c^2}+{2 \over 5}\left(E\over m_{\rm e}c^2\right)^2\right]^{1/2}.
 \label{Compton}
\end{equation}
With these assumptions, the standard radiation transfer equation is solved iteratively in the range of Thomson optical depth of $10^{-4} < \tau < 10$. At each point within the slab, the population levels, temperature, opacity and emissivity are calculated using {\tt XSTAR}. As the radiation penetrates within the slab, the continuum is modified by both scattering and absorption (followed by re-emission at different energy). At some depth, recombination leads to a rapid cooling and the temperature falls to $\sim 10^5$\,K. Hence, the temperature structure of the slab is approximately constant followed by a sudden drop after a certain optical depth. The ionization parameter has two major effects on the temperature structure. For higher ionization parameter, first, the temperature of the illuminated region is high, and second, the radiation penetrates deeper the slab.

Let us consider our specific cases of $\log_{10} \xi= 2.7$ and 3.3. As we see in Fig.~\ref{ionization_broadening}(b), the latter case leads to a much broader line than the former. The main differences between these two cases are as follows. (i) Difference in ionization states: At these ionization levels, there is no Fe K$\beta$ emission. For $\log_{10} \xi=2.7$, the emission spectrum is rich with many Fe K$\alpha$ lines in the 6.4--6.7\,keV band, while for $\log_{10} \xi=3.3$, the line emission is centred at 6.7\,keV and 6.9\,keV of the Fe K$\alpha$ emission from highly ionized Fe. (ii) For the former case, the illuminated region of the gas attains a temperature of $kT\lesssim 1$\,keV, in which case Fe K$\alpha$ photons are only down-scattered when leaving the disc. In the latter case, the temperature is much higher, $kT \gtrsim 10$\,keV. At this temperature, the probability of up and down-scattering becomes comparable, see equation (\ref{Compton}); hence, the line is symmetrically broadened. (iii) In the former case, the photoelectric opacity dominates over scattering in the 10--$10^4$\,eV range, while the opposite occurs for the latter case. Hence, the absorption dominates in the lower ionization case, leading to decreased flux level at both the red and blue wing of the line complex. (iv) In the former case, the illuminating radiation fully ionizes the Fe only down to $\tau\sim 0.2$, while in the latter case, the ionization continues till $\tau\sim 1$. Consequently, for higher ionization, the line complex is produced deeper inside the slab, and the line photons encounter more Compton scattering while coming out, which leads to a broader profile (e.g., \citealt{st80}).

We have also tested whether the data of obs.\ 5--7 are also consistent with low values of $\xi$ and $r_{\rm in}$. For this purpose, we have modified the case 2(i) by imposing low values of $\xi$. In particular, we fix $\log_{10} \xi=2.4$ (found for obs.\ 1) for all observations. We then obtain a much worse fit with $\chi^2$/dof = 1119/844, i.e., $\Delta \chi^2=+348$ for one less free parameter. We still find the disc to be truncated, with $r_{\rm in}$ in the range 20--600. For obs.\ 7, we obtain $r_{\rm in}=20.0\pm 3.3$.

\subsection{Comparison with \xte}
\label{xte}

\begin{figure}\centering{
\includegraphics[width=\columnwidth]{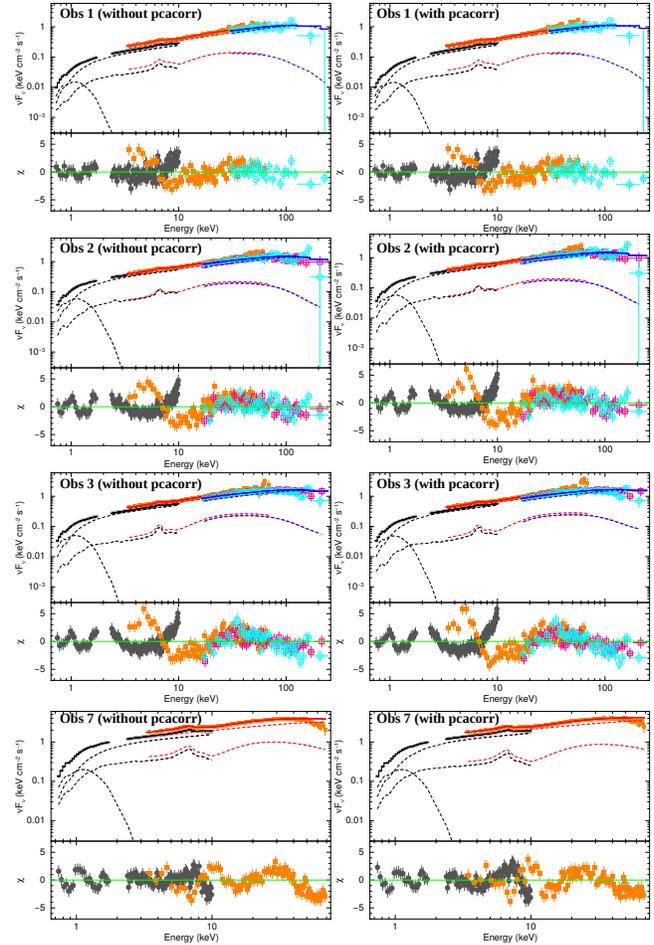} }
\caption{Broad-band spectral fits with \xmm\ and \xte\ for the observations with available simultaneous \xte\ data (obs.\ 1, 2, 3 and 7). The symbols used are: \xmm\ EPIC-pn: gray filled circles, \xte\/ PCA: orange filled squares, \xte\/ HEXTE: cluster A --- magenta open squares, cluster B --- light blue open circles. The data are fitted with the model 1(i), with model components shown by dashed lines and the full model shown by solid lines. The left and right panels are without and with the {\tt pcacorr} correction, respectively. The PCA data are used for all cases, while the HEXTE data is used whenever good quality data are available. 
}
\label{xmm_xte}
\end{figure}

Although our work is devoted to the analysis of the \xmm\/ EPIC-pn data, we have used the simultaneous \xte\/ PCA and HEXTE data to calculate the bolometric fluxes given in Table \ref{tobs}, except for the obs.\ 4--6, where we use the spectrum obtained from the \swift\/ Burst Alert Telescope (BAT; \citealt{segreto10}). For the HEXTE data, we choose one or both of the clusters A and B according to their observed count rate and the available background. We use both the clusters for obs.\ 2 and 3, while cluster B is used for obs.\ 1. For obs.\ 7, while the source data was obtained by cluster A, the background is not available. Hence, we could not use the HEXTE data for this case. Then we fit all the spectra using the model of Case 1(i). As a wider energy range is available, the electron temperature of the {\tt nthcomp} $kT_{\rm e}$ and the cutoff energy of the {\tt relxill} $E_{\rm c}$ are initially made free. However, we found a large difference between $kT_{\rm e}$ and $E_{\rm c}$. Hence, we put a constraint as $E_{\rm c} = 1.5 kT_{\rm e}$ except for obs.\ 7 for which $E_{\rm c} = 3 kT_{\rm e}$ is assumed.

Fig.~\ref{xmm_xte} shows the resulting fits, which clearly show the mismatch of the calibration between \xmm\/ EPIC-pn and \xte\/ PCA data in the Fe K region. For the PCA data, we have also checked the effect of applying the calibration correction {\tt pcacorr} of \citet{garcia14b}, which we found to be negligible, as illustrated in Fig.~\ref{xmm_xte}. The systematic error applied to the PCA data is 0.1\% with the {\tt pcacorr} correction and 0.5\% without the correction. We have checked that the observation times overlap well; hence, the discrepancy cannot come from the difference it. This apparent calibration difference between the EPIC-pn and PCA may possibly explain the differences between our results and those of \citet{garcia15}, see Appendix \ref{comparison}.

\subsection{Hysteresis in the reflection properties}
\label{hysteresis}

As we see in Fig.\ \ref{ionization_broadening}(a), the profiles of the reflection spectra correlate with the spectral slope rather than with the flux. Namely, the obs.\ 7 has the flux four times that of the obs.\ 5--6, and the obs.\ 2, 3 have the flux twice that of the obs.\ 1, yet they show very similar spectra within each trio. This indicates that the basic properties of a state of \source\ are determined by the spectral hardness rather than the flux. In the framework of the truncated disc paradigm, this would imply the disc truncation radius, which determines the spectral properties, depending on the history of a given outburst. For example, obs.\ 4--6, which took place during the declining phase of `failed' (i.e., without going to the soft state) outburst, have apparently similar truncation radius as the obs.\ 7, taken during the rise of an outburst. In the case of ionization affecting the line width (Section \ref{ionization}), our results imply that the ratio of the flux to the disc density is constant for a given spectral hardness. Our results indicate that both disc truncation and ionization are important; the reflection fraction and the ionization parameter increase and the disc inner radius decreases with the increasing spectral softness.

This is another manifestation of hysteresis, well known to occur in \source, in which the hard-to-soft transition usually occurs at a much higher $L$ than the reverse transition. Hysteresis is a feature of systems in which there are two stable states for the same controlling parameter, the accretion rate in our case. Once a system goes to a given state it stays in it as long as it is physically possible, which leads to the value of the controlling parameter at the transition depending on its direction.

\subsection{Caveats}
\label{caveats}

We have done our best to determine the actual disc truncation radius in the hard state. In our best models, this radius decreases with the softening slope of the X-ray continuum, but it remains high through the hard state, with $r_{\rm in}$ changing from a few hundred to a few tens as $\Gamma$ changes from 1.5 to 1.7. 

Still, these results are based on currently available models, which still have limitations. In particular, the ionization structure in {\tt xillver} is calculated for a fixed density, and irradiating photons are assumed to hit the disc at an angle of $45\degr$. Relaxing those assumptions may possibly lead to some changes of the results. We also note that the definition of the reflection fraction in {\tt xillver} and {\tt relxill} (v0.2h) is phenomenological, which issue we discuss in Appendix \ref{reflection}. However, this does not affect our determination of other parameters of the system.

Furthermore, \citet*{nzs16} have pointed out that the direct emission is not redshifted in the {\tt relxill} and {\tt relxilllp} models, and photons reflected by the disc and returning to it due to light bending are not taken into account. However, our results show only the presence of only moderate relativistic effects in \source; thus, those issues do not affect our results. 

Then, we have found out that assuming the disc inner radius determined from the reflection and that from the disc continuum are the same leads to an apparently unphysical result, of $r_{\rm in}$ decreasing for harder spectra. This is apparently related to the fact that the disc blackbody model ignores the effect of its irradiation by the X-rays emitted above $\sim$ 1 keV, which, in the hard state, have the luminosity content an order of magnitude higher than that of the disc itself, see \citet{gierlinski08,gierlinski09} and KDD14. For example, \citet{rykoff07} found that an outer disc region is strongly irradiated, emitting flux in the far UV range significantly exceeding that expected from a standard disc. But the effect of irradiation of the inner disc in X-rays is usually ignored, though the recent study by \citet{vincent16} takes it into account. We have tried to circumvent those problems by constraining our fits to energies $\geq 2.35$ keV, where the disc contribution is weak. Thus, we do not expect that this would affect our final results for $r_{\rm in}$ from reflection. 

We also note that the normalization of the disc blackbody model is $\propto r_{\rm in}^2$. For weak disc components, $r_{\rm in}$ will be low. In the vicinity of the BH, the natural scale length is $R_{\rm g}$, and clouds or a small inner disc may be formed with that characteristic size. This may explain results of papers claiming the disc extending to the ISCO based on disc fitting. In our work, we have found that we do not require the presence of clouds or a small inner disc, but this may be simply due to the limited statistics of the data. 

We have also not explored the effect of using more complex continua, e.g., two Comptonization components, see \citet{yamada13} and KDD14. Such complex continua are required by the physics of the accretion. That complexity may be responsible for the soft components, fitted by us and others as discs.

\section{Conclusions}
\label{conclusions}

We have analysed the seven available X-ray observations of \source\ in the hard spectral state by \xmm\/. We have jointly fitted the spectral data by Comptonization and the currently best available Compton reflection code \citep{gk10,dauser10}, {\tt relxill}. In the fitting, we have assumed the parameters of the interstellar absorption, and the reflector abundances and inclination to be the same for all the data. We have studied a large array of spectral models, in particular, we have considered in detail a possible contribution from a standard blackbody accretion disc. Also, we have reviewed the parameters of the \source, and, in particular, its transient behaviour. Our main results are as follows.

In Section \ref{gx}, we have compared the estimates of the average accretion rate of \source\ by \citet{z04} and \citet{cfd12}, and found them to be in agreement. On the other hand, its theoretical estimates based on the stripped giant model by \citet{munoz08} yield much lower values, indicating a need to revise these calculations and a tentative character of the current estimates of the properties of the donor.

We have tested whether the inner radius of the blackbody disc in the spectral fits can be set equal to that of the reflector. In order to be able to do it, we have modified the {\tt diskbbb} model (Section \ref{diskbb}). However, we have found (Section \ref{full}) that this leads to an unphysical behaviour of the disc truncation radius, which increases with the spectral softness. This increase is not consistent with either the truncated disc model, in which the truncation radius decreases to the ISCO when approaching the soft state, or the disc extending down to the ISCO during all the observations. This implies that the soft X-ray component is not a standard blackbody disc. This is fully consistent with the presence of strong disc irradiation by the hard X-rays \citep{gierlinski08,vincent16}, which dominate the total emission by a large factor in the hard state. We thus treat the soft component phenomenologically in our final models.

We have considered a large range of models, testing among others the effects of the chosen energy range, of adding an unblurred reflection component, and assuming a lamppost geometry. Surprisingly, we have found the effects of relativistic broadening to be relatively weak in all cases. In the coronal models, we find the inner radius to be large, $r_{\rm in}\gg 1$. In the lamppost model, the inner radius is unconstrained, but when we fix it to the ISCO, we find the height of the source to be large, which also implies a weak relativistic broadening. Still, we find the inner radius in the coronal models to correlate with the X-ray hardness ratio, which is consistent with the truncation radius decreasing from being large in quiescence to the ISCO in the soft state. Our large values of the disc truncation radius agree well with those determined by \citet{demarco15}, based on X-ray reverberation time lags, and obtained using the observations 1, 2--3 and 7. 

We have found the disc inclination of $\simeq 40\degr$--$60\degr$, implying $M\gtrsim 8\msun$ for the case of the disc and binary axes being aligned. The fitted Fe abundance has been found to be approximately solar. 

We have also found a strong effect of the disc ionization. Its degree anti-correlates with the hardness, with the ionization parameter becoming very large in our softest states. This leads to strong spectral broadening due to Compton scattering of the reflected photons in the disc, and accounts for the relatively large truncation radii even in the softest studied states. We have discussed the physics of this effect in Section \ref{ionization}. If we constrain the ionization to be weak, the fit strongly worsens but still $r_{\rm in}\gg r_{\rm ISCO}$. 

Although our study is devoted primarily to the \xmm\/ data, we have also analysed (Section \ref{xte}) the simultaneous \xte\/ data available for obs.\ 1--3 and 7, in order to determine the bolometric flux. At this occasion, we have found pronounced differences in the calibration of the EPIC-pn and the PCA in the Fe K region, see Fig.\ \ref{xmm_xte}. This disagreement persists even when the correction proposed by \citet{garcia14b} is applied. This may explain the differences of our results with respect to those of \citet{garcia15}, who fitted the PCA data only.

We have found that the properties of the reflection component are mostly determined by the spectral hardness, with both the relative strength and the broadening of the reflection spectrum decreasing with the hardness. On the other hand, these properties are not determined by either the 1--10 keV or the bolometric flux. This appears to be one more manifestation of the hysteresis, well known to be present in \source\ (Section \ref{hysteresis}).

In Appendix \ref{comparison}, we have compared our results with those of other studies. We find the differences accounted for by the effect of pileup when fitting the EPIC-MOS data, and by using less advanced spectral models. Still, we have found our results generally consistent with those of \citet{plant14b,plant15} and with a number of fits in \citet{furst15}.

In Appendix \ref{reflection}, we have discussed in detail the issue of the definition of the fractional reflection strength. We have found that the phenomenological definition based on the 20--40 keV flux of the reflected photons used in {\tt xillver} and {\tt relxill} v0.2h leads to some unphysical effects. We advocate the use of the definition based on the ratio of the flux of photons directed in the local frame towards the disc and away from it.

\section*{Acknowledgments}
We thank Bei You and Andrzej Nied{\'z}wiecki for valuable discussions, Javier Garc{\'{\i}}a for help with the reflection codes, and Alberto Segreto for providing us with the BAT spectrum for the observations 4--6. This research has been supported in part by the Polish NCN grants 2012/04/M/ST9/00780 and 2013/10/M/ST9/00729, and it has made use of data obtained through the HEASARC Online Service, provided by the NASA/GSFC, in support of NASA High Energy Astrophysics Programs.

\appendix

\section{Comparison with previous spectral results}
\label{comparison}

We present here a detailed comparison of our results with those of other authors using the same data, as well as with studies of the hard-state data of \source\ obtained by other satellites. We consider the spectral state to be hard for $\Gamma\lesssim 1.9$.

\citet{miller06} analysed the obs.\ 2 and 3 using the EPIC-MOS data, and the angle-averaged reflection model {\tt cdid} of \citet{ballantyne01} convolved with the relativistic {\tt kdblur} (angle-dependent) model \citep{laor91}. They obtained $r_{\rm in}\simeq 5.0\pm 0.5$ and $i\simeq 20^{+5}_{-10}\degr$. \citet{reis08} analysed the same data using the {\tt reflionx} reflection model (angle-averaged; \citealt{ross_fabian05}) and the relativistic {\tt kdblur} model. They obtained $r_{\rm in}= 2.04^{+0.07}_{-0.02}$ and $i\simeq 20\degr$. \citet{miller08} analysed the same data using the {\tt cdid} reflection model and the relativistic {\tt kerrconv} model \citep{br06} assuming $r_{\rm in}=r_{\rm ISCO}$. They obtained $a=0.93\pm 0.1$ and $i\simeq 19\degr\pm 1\degr$ (including also other data in the very high and intermediate states). Then \citet{reis10} analysed again the same data using the {\tt laor} model and obtained $r_{\rm in}= 2.4^{+0.3}_{-0.5}$ and $i=10^{+17}\degr$. In all cases, the difference with respect to our results is mostly due to using the MOS data by those authors. 

\citet{reis10} also fitted the soft component in those data as a blackbody disc with the zero-stress inner boundary condition using the {\tt ezdiskbb} model \citep{zimmerman05}, and obtained a similar result of $r_{\rm in}= 3.5^{+4.4}_{-2.2}$. As we found out in this work, the behaviour of the soft component is not compatible with being a disc blackbody. Indeed, \citet{reis10} found, see their fig.\ 5, that $r_{\rm in}$ increases with the luminosity in their sample of BHBXBs, which is not compatible with the transition to the soft state (where $r_{\rm in}=r_{\rm ISCO}$) at the highest $L$ of the hard state.

\citet{Done_Diaz_2010} have analysed the \xmm\/ obs.\ 2 and 3, and have shown that the claim of \citet{miller06} of the detection of the disc down to the ISCO is due to an instrumental effect, namely pile-up in the MOS data, as we stated in Section \ref{intro}. The fits of \citet{Done_Diaz_2010} to the EPIC-pn data yield results consistent with ours.

KDD14 analysed the obs.\ 1, 2, 3 and 7 (denoted by them as 1, 3, 2, 4, respectively, ordered by the increasing flux), using the EPIC-pn data. Their fig.\ 2 shows that the strength and width of the Fe K feature significantly increases with the increasing flux, which we also find in our data, in which the increasing flux corresponds approximately to the increasing $\Gamma$. They have shown that the EPIC-pn calibration has some problems, in particular, an apparently instrumental dip at $\simeq$9.4 keV, and a significant difference in the fitted spectral slopes between the EPIC-pn and the \xte\/ PCA. For \source, they have shown the disagreement between the values of $r_{\rm in}$ obtained from the disc and reflection components, as well as the sensitivity of these values to the assumed model, in particular the effect of using two Comptonization components, see their fig.\ 8. 

\citet{plant15} have fitted the \xmm\/ obs.\ 1, 2, 3 and 7 together with simultaneous \xte\/ data, and two \suzaku\/ observations. They used a previous, slightly different, version of {\tt relxill}, i.e., {\tt relconv} \citep{dauser10} convolved with {\tt xillver}. They fitted the \xmm\/ data at both $\geq$4 and $\geq$1.3 keV ranges. They obtained very similar results to ours, with $r_{\rm in}$ decreasing from $>100$ for obs.\ 1--3 to $\sim$70 for the obs.\ 7. They fitted inclination is also similar to ours, $i=42^{+11}_{-6}\degr$. They also used {\tt reflionx} instead of {\tt xillver}, obtaining similar results. Overall, their results are fully consistent with ours. They also have shown that fitting data with only the relativistic line profile and neglecting the associated reflection continuum (as, e.g., in \citealt{reis10}) leads to underestimating the actual truncation radius by about an order of magnitude (see their fig.\ 9).

\citet{plant14b} have fitted the \xmm\/ obs.\ 4, 5, 6 in the 0.4--10 keV range, using {\tt relxill} and {\tt diskbb}. They found somewhat different results from ours, with $r_{\rm in}\sim$20--30 for both reflection and disc, and $i\simeq 30\degr$. The differences are accounted for by the different bandpass fitted, the $A_{\rm Fe}$ fixed at 1, and some other minor differences. Still, their results support the truncated disc model.  

\citet{furst15} analysed five simultaneous \nustar\/ and \swift\/ XRT observations during 2013 Aug.--Oct., in the 0.8--80 keV range. Their main finding appears that in order to obtain good fits with an expected Fe abundance, $A_{\rm Fe}\lesssim 2$, they require certain complexity of the continuum, namely the local spectra close to the BH (which dominate the reflection due to light bending) to be harder than those farther away (which dominate the direct spectrum, especially at low energies). This spectral hardening with the decreasing radius, i.e., following the accretion flow, is fully consistent with a number of other findings, e.g., the presence of hard lags \citep{kotov01}, spectral variability requiring multiple Compton components, softer at larger radii \citep{yamada13}, and the concave shape of the $\sim$1--10 keV continuum of BHXRBs, e.g., \citet{gierlinski97}, \citet{frontera01}, \citet{disalvo01}.

The authors state that most of their models require $r_{\rm in}$ close to the ISCO. Indeed, some of their complex models, e.g., the 'lamppost' model with two sources along the BH rotation axis at different heights give $r_{\rm in}\sim 3$--10 for the first four observations. This is also the case for some models with $A_{\rm Fe}\gtrsim 5$. However, many of their models, including some with the lowest reduced $\chi^2$, have the best-fit values of $r_{\rm in}$ between several tens and $>10^2$. Thus, their results allow either low or high values of $r_{\rm in}$. They note that the \xmm\/ obs.\ 4--6 were taken between their obs.\ IV and V. \citet{furst15} fitted them with their model M2-q3, obtaining results similar to those for the \nustar-\swift\/ observations, i.e., $r_{\rm in}\sim 10^2$, compatible with the results presented here.

\citet{tomsick08} fitted two \swift\/ XRT/\xte\/ PCA hard-state observations during 2007 May--June. They modelled reflection using the {\tt pexriv} model \citep{mz95} with relativistic broadening modelled with the {\tt kdblur}  and a separate {\tt laor} Fe line. They obtained $r_{\rm in}\simeq 2$--5. We note that the ionization treatment of {\tt pexriv} is highly approximate, using the ionization calculations of \citet{done92}. Also, the fits were done assuming $i=20\degr$. Thus, we consider that evidence for the strongly relativistic line as uncertain. Also, they found that their {\tt diskbb} component for the first spectrum implies a significantly higher $r_{\rm in}$.

\citet{tomsick09} have fitted a \suzaku\/ and \xte\/ observation in the hard state on 2008 September during an outburst decline, for which the 1--100 keV unabsorbed flux was $2.4\times 10^{-10}$ erg cm$^{-2}$ s$^{-1}$, which corresponds to 0.13 per cent of $L_{\rm E}$ at 8 kpc and $10\msun$. They used a simplified model of a power law and a broadened Fe K line using the {\tt laor} \citep{laor91} model (i.e., neglecting reflection). They found a highly truncated disc, with $r_{\rm in}>175$ for $i\geq 30\degr$.  

\citet{shidatsu11} analysed three \suzaku\/ observations in the hard state in 2009 March using {\tt pexriv} and {\tt diskline} \citep{fabian89}. We note that this is a rather approximate model; as we noted above, the former is not accurate, and the latter Schwarzschild metric model neglects the light bending. The unabsorbed flux of the observation in the 0.5--300 keV range was $\simeq 5\times 10^{-9}$ erg cm$^{-2}$ s$^{-1}$, which corresponds to $\simeq$3 per cent of $L_{\rm E}$. They obtained $r_{\rm in}=13.3^{+6.4}_{-6.0}$ and $i=46\pm 8\degr$. This is compatible with a truncated disc, though their radius is lower than ours. They also re-analysed the 2008 Sept.\ observation of \citet{tomsick09} and confirmed their result, obtaining $r_{\rm in}>190$ for the fixed $i=50\degr$.

\citet{allured13} analysed seven \swift\/ XRT and \xte\/ observations in the hard state between 2007 May to 2010 March with the range of $L$ between 0.4 and 5 per cent of $L_{\rm E}$. They modelled reflection by {\tt kdblur} and {\tt reflionx}, and disc emission by {\tt diskbb}. They assumed either $i=20\degr$ or $50\degr$. The reflection fits gave $r_{\rm in}$ in the range from $\simeq$3 to $\sim$40, and $\simeq$10--100 at these two inclinations, respectively, with no clear dependence on $L$. The disc fits gave $r_{\rm in}\sim 10$ for most of the data.

\citet{petrucci14} analysed five \suzaku\/ observations in the hard state in 2011 February--March. The unabsorbed 0.7--70 keV fluxes declined from $\simeq 1.6$ to $0.24\times 10^{-9}$ erg cm$^{-2}$ s$^{-1}$, which corresponds to $\simeq$0.8--0.1 per cent of $L_{\rm E}$. Note that the used range does not cover the energy range with most of flux in the hard state, around $\sim$100 keV, so the actual fluxes can be larger by a factor of $\gtrsim 2$. They modelled the reflection by {\tt kdblur} and {\tt reflionx}. They found disc inner radii in the range of $r_{\rm in}>11$ to $>70$ assuming a power-law incident spectrum, and somewhat less, $>6$ to $>70$ for the continuum modelled as Comptonization by {\tt eqpair} \citep{coppi99}. They fixed the inclination used by {\tt kdblur} at $20\degr$, while the reflection spectrum used by them was angle-averaged.

\citet{garcia15} analysed the \xte\/ PCA data in hard state. They found evidence for the presence of an unblurred reflection component, and consequently added an additional {\tt xillver} component. They did not include a {\tt diskbb} model. They found that the values of $r_{\rm in}$ are close to the ISCO. The other parameters found by them (e.g., $i=48\degr\pm 1\degr$) have usual values except for the $A_{\rm Fe}$, which was found very high, $\simeq 5\pm 1$. We consider this to be a major caveat for their results. Furthermore, the profiles of the Fe K region in fits to simultaneous \xte\/ PCA and \xmm\/ EPIC-pn data disagree systematically, even including the PCA calibration correction of \citet{garcia14b}, see Section \ref{xte}.

\section{The reflection fraction}
\label{reflection}

The issue of how to measure the relative strength of reflection is quite complex. In the non-relativistic (i.e., neglecting relativistic effects on the propagation of the reflected radiation), angle-dependent, models of {\tt pexrav} and {\tt pexriv} (and the associated convolution models, {\tt reflect} and {\tt ireflect}; \citealt{mz95}), the reflection fraction, ${\cal R}$, is defined as the ratio of the flux directed towards the observer to that toward the slab. Thus, ${\cal R}=1$ corresponds to isotropic, optically-thin emission reflected from an infinite slab. Then ${\cal R}$ can be roughly identified with the solid angle, $\Omega$, subtended by the reflector, ${\cal R}\simeq \Omega/2\upi$. Naturally, ${\cal R}\neq 1$ can also correspond to either the direct emission being not isotropic or a part of the primary source being obscured (e.g., in Seyfert-2 galaxies). Also, some of the reflected radiation can be up-scattered by a hot corona, and thus removed from the reflected spectrum.

The above definitions involve photons hitting the reflecting slab, not those reflected. The number of the reflected photons can be $\ll$ than that hitting the slab, as the process involves both scattering and absorption. In general, the albedo integrated over a typical power law spectrum with a high energy cutoff at $\sim$100 keV is $\ll 1$, but its value strongly depends on the ionization state. The above distinction is often not taken into account, with some papers giving the ratio of the reflected to direct fluxes as the reflection fraction. 

Some models, e.g., {\tt reflionx} \citep{ross_fabian05,ross_fabian07}, do not provide a reflection fraction at all, and instead normalize the reflection strength by the ionization parameter, proportional to the flux incident on the reflector. Then, however, it is unclear what geometry a given model corresponds to. Comparing the direct and reflected fluxes can give a rough estimate, which, however, may be quite inaccurate given the unknown albedo (dependent on the spectral shape and the degree of ionization).

\begin{figure}\centering{
\includegraphics[angle=0,width=6.8cm]{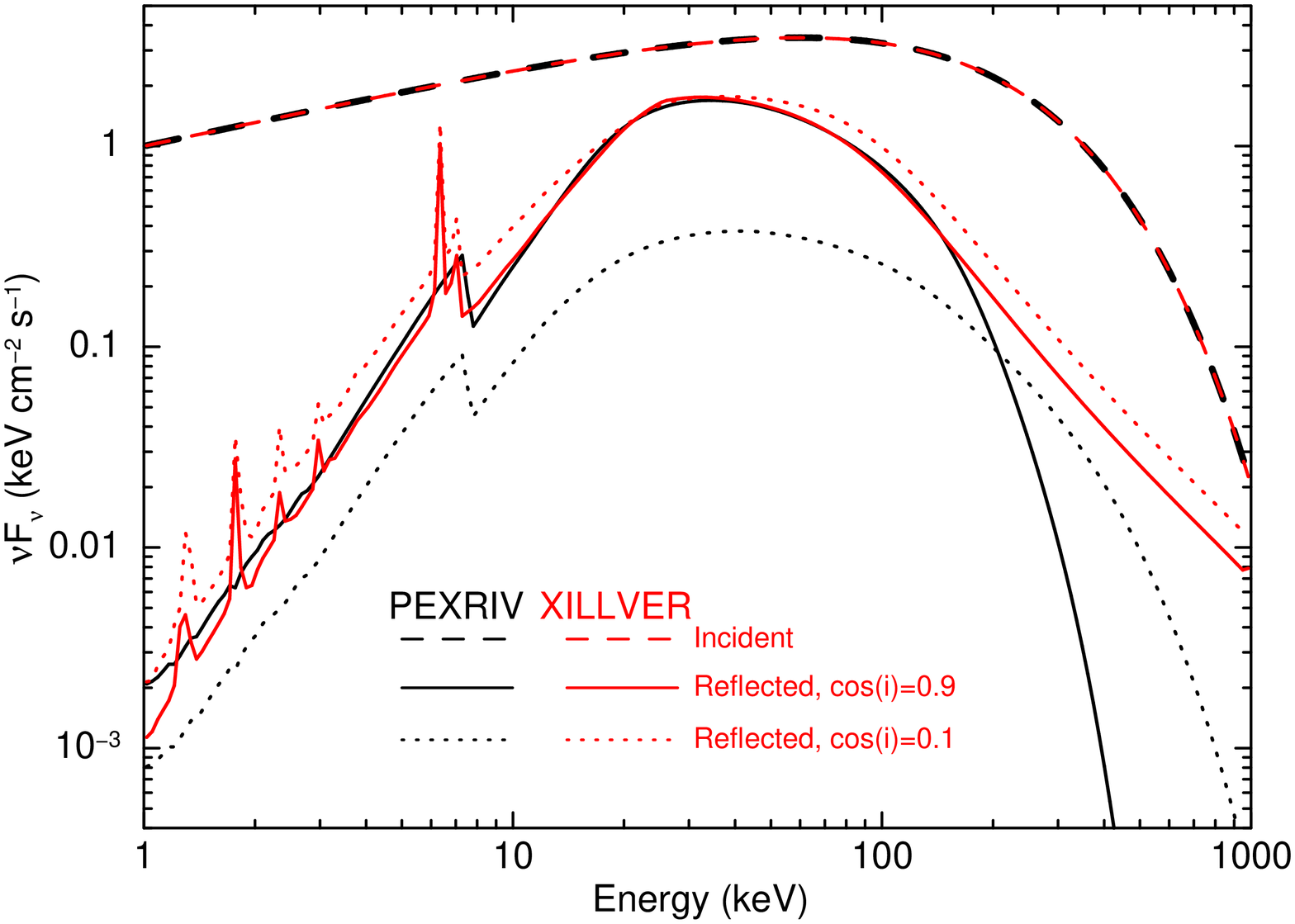} }
\caption{Comparison of the reflection spectra from {\tt xillver} and {\tt pexriv} for $\cos i = 0.1$ and 0.9, at $\Gamma = 1.6$, $E_{\rm c} = 150$ keV, $\xi=10$, ${\cal R}=0.5$. For a given geometry, the reflected spectrum should decrease with the increasing viewing angle (and become null at $i=90\degr$). In contrast, {\tt xillver} at a fixed ${\cal R}$ normalizes the reflected spectra to a constant flux in the 20--40 keV range, which leads to the reflected spectra almost independent of the inclination, which is not physical. We also note that the shape of the reflected spectra from {\tt xillver} at a few tens of keV significantly differ from those of {\tt pexriv}, which is due to the non-relativistic treatment of Compton scattering by the former, see equation (\ref{Compton}), which fails at high energies.
}
\label{xillver_pexriv}
\end{figure}

\begin{figure}\centering{
\includegraphics[angle=0,width=6.8cm]{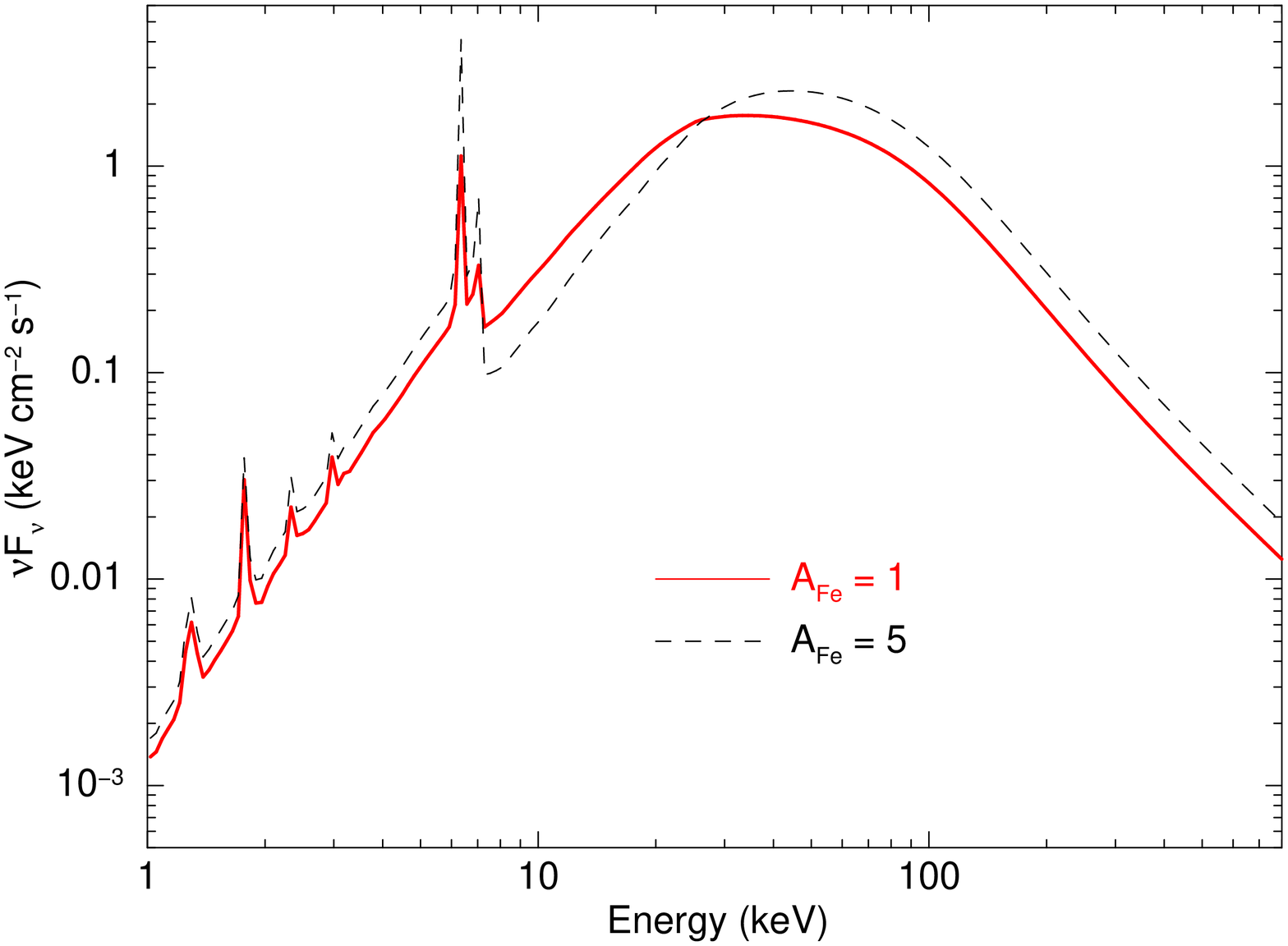} }
\caption{The effect of the Fe abundance on {\tt xillver} reflection spectra, for $A_{\rm Fe}=1$ and 5, at $\Gamma = 1.6$, $E_{\rm c} = 150$ keV, $\xi=10$, 
$i = 60 \degr$. Since the irradiating spectrum is constant, and the Fe abundance plays no role for photons at high energies, the reflected spectra should be identical at $E\gtrsim 50$ keV, while they are not owing to the code normalizing the reflected spectra to a constant flux in the 20--40 keV range. This leads to a further unphysical effect that the reflected component below the Fe K$\alpha$ line increases with the increasing $A_{\rm Fe}$, while it should decrease, due to the opacity increased by the additional Fe.
}
\label{AFe}
\end{figure}

The models we use in this work, {\tt xillver} and {\tt relxill} v0.2h, define ${\cal R}$ by the energy flux ratio between the reflected and direct observed spectra in the 20--40 keV range regardless of the inclination. This, unfortunately, leads to quite significant inaccuracies. We show some examples (for {\tt xillver}) in Figs.\ \ref{xillver_pexriv} and \ref{AFe}. Fig.\ \ref{xillver_pexriv} shows that {\tt xillver} yields an approximately correct amplitude of the reflected spectra for an inclination close to face-on, but overestimates it at large viewing angles. Also, we note that the shape of the high-energy cutoff in the reflected spectra becomes incorrect above $\sim$100 keV, due to the scattering treatment of equation (\ref{Compton}) being non-relativistic. Fig.\ \ref{AFe} shows that the used range of 20--40 keV is also sensitive to the Fe abundance, and its change leads to a change of the high-energy reflected continuum, which, as dominated by scattering, should be independent of it.

In our opinion, the most physical way of defining ${\cal R}$ in the non-relativistic case is that of {\tt pexrav/pexriv}. Namely, ${\cal R}$ should be defined as the ratio of the flux directed towards the observer to that toward the disc. This appears straightforward to implement in {\tt xillver}. Following our discussion, \citet{dauser16} have implemented that definition in {\tt xillver} v0.4a.

On the other hand, further complexity appears when relativistic effects on the photon propagation are taken into account, due to both photons no longer travelling on straight lines and some of them crossing the BH horizon. These effects can be quite extreme if most of the emission is from close vicinity of a rotating BH. For a given static reflection model, they can be taken into account, e.g., using the {\tt relconv} model \citep{dauser10}. In the case of {\tt xillver}, the relativistic effects are taken into account by {\tt relxill} (but see \citealt{nzs16}). Still, {\tt relxill} defines the reflection strength by the flux ratio in the 20--40 keV energy range. This creates the same problems as shown in Figs.\ \ref{xillver_pexriv} and \ref{AFe}. 

Then, \citet{dauser14} defined the reflection fraction as the ratio of the photon flux hitting the disc to that for photons escaping to infinity. They show that this ratio can reach very large values, up to $\sim$30, for the maximally rotating BH surrounded by a disc extending down to the ISCO. However, this definition disagrees with that in terms of the 20--40 keV flux ratio, and the disagreement can be very strong due to the beaming of both direct and reflection radiation toward the equatorial plane of a rotating BH. For that reason, the definition of \citet{dauser14} has been implemented in {\tt relxill} v0.4a \citep{dauser16}.

On the other hand, our proposal is to directly extend the non-relativistic definition used by {\tt pexrav}. The reflection fraction, ${\cal R}$ would be then defined as the ratio of the flux directed toward the disc to that away from it {\it in the local frame}. In coronal models, the irradiating flux is usually assumed to be power-law function of the radius. Then, ${\cal R}=1$ would correspond to the local-frame emission being isotropic at each radius. If needed, it would be easy to calculate the fractional contribution of reflection in any given photon-energy band by integrating over the model components.

In the lamppost model, ${\cal R}=1$ would correspond to the case with an isotropic point source on axis of the BH, and ${\cal R}\neq 1$ would scale that theoretical reflection spectrum. E.g., if either ${\cal R}\gg 1$ or ${\cal R}\ll 1$ were found in fitting data, it would indicate an inadequacy of the assumed model.

\label{lastpage}

\end{document}